\documentclass[preprint,showpacs,preprintnumbers,amsmath,amssymb]{revtex4}

\usepackage{graphicx}\graphicspath{./graph/}%
\usepackage{dcolumn}% Align table columns on decimal point
\usepackage{bm}% bold math
\usepackage{color}
\usepackage{verbatim}
\usepackage{amssymb}
\usepackage{amsfonts}
\usepackage{latexsym}
\usepackage{epstopdf}
\usepackage{url}
\usepackage{float}
\usepackage{placeins}
\definecolor{red}{rgb}{1,0,0}
\definecolor{green}{rgb}{0,1,0}
\definecolor{blue}{rgb}{0,0,1}

\newcommand{\be}{\begin{equation}}
\newcommand{\ee}{\end{equation}}
\newcommand{\bea}{\begin{eqnarray}}
\newcommand{\eea}{\end{eqnarray}}
\newcommand{\bdm}{\begin{displaymath}}
\newcommand{\edm}{\end{displaymath}}

\newcommand\ptl{\partial}

\newcommand\imp{{\rm Im}}
\newcommand\rep{{\rm Re}}

\newcommand\td{\tilde}
\begin{document}
%\preprint{APS/123-QED}
\title{Equilibrium of an Arbitrary Bunch Train in the\\ Presence of
Multiple Resonator Wake Fields}
\author{Robert Warnock }
\email{warnock@slac.stanford.edu}
\affiliation{SLAC National Accelerator Laboratory, Stanford University, Menlo Park, CA 94025, USA}
\affiliation{Department of Mathematics and Statistics, University of New Mexico, Albuquerque, NM 87131, USA}

\begin{abstract}
A higher harmonic cavity (HHC),  used to cause bunch lengthening for an increase in the Touschek lifetime, is a feature of several  fourth generation synchrotron light sources. The desired bunch lengthening is complicated by the presence of required gaps in the bunch train. In a recent paper the author and Venturini studied the effect of various fill patterns by
calculating the charge densities in the equilibrium state, through coupled Ha\"issinski equations. We assumed that the only
collective force was from the beam loading (wake field) of the harmonic cavity in its lowest mode. The present paper improves the notation and
organization of the equations so as to allow an easy inclusion of multiple
resonator wake fields. This allows one to study the effects of beam loading of the main accelerating cavity,  higher order modes of the cavities,
and short range geometric wakes represented by low-$Q$ resonators. As an example these effects are explored for ALS-U. The compensation of the induced voltage in the main cavity, achieved in practice by a feedback system, is modeled by adjustment of the generator voltage
through a new iterative scheme. Except in the case of a complete fill, the compensated main cavity beam loading has a substantial effect on the bunch profiles and the Touschek lifetimes. A $Q=6$ resonator, approximating the effect of a realistic short range wake, is also
consequential for the bunch forms.
\end{abstract}
\maketitle
\section{Introduction}
This is a sequel to Ref.\cite{prabI}, in which we explored the action of a higher harmonic cavity (HHC), a standard component of 4th  generation synchrotron light sources, employed to lengthen the bunch and reduce the effect of Touschek scattering.   In that work we introduced an effective scheme to compute the equilibrium state of charge densities in an arbitrary bunch train. The train is allowed to have arbitrary gaps and  bunch charges. We chose the simplest possible physical model, in which the only induced voltage (wake field) is due to the lowest mode of the HHC.
We write $V_{r3}$ for this voltage, the notation designating ``resonator, 3rd harmonic". We recognized, however, that excitation of the main accelerating cavity (MC) by the bunch train produces an induced voltage $V_{r1}$ of comparable magnitude, the effect described as {\it beam loading}. Our excuse for omitting $V_{r1}$ was that in practice it is largely cancelled by adjusting the rf generator voltage $V_g$ through a feedback system. The sum of $V_{r1}$ and $V_g$ should closely approximate $V_{rf}$, the desired accelerating voltage.

In real machines there are always  gaps in the bunch train, and that leads to varying bunch profiles and centroid displacements along the train.
At first sight this would suggest that $V_{r1}$ would be different for different bunches, so that compensation could only be partial, perhaps only manifest in some average sense.  On the contrary,  we shall calculate an equilibrium state in which the compensation is essentially perfect for all bunches. {\it This happens by an adjustment of  the charge densities of all bunches, to new forms that sometimes differ substantially from those without the MC.} The adjustment is achieved automatically through a new algorithm presented here. This iterative procedure minimizes a mean square deviation of
$V_{r1}+V_g$ from $V_{rf}$, summed over all bunches, as a function of two generator parameters, which are equivalent to amplitude and phase.

It is not clear that this is a faithful model of the feedback mechanism, which  could  conceivably amount to a weaker constraint on the bunch profiles. Nevertheless, this study clarifies the mathematical structure of the problem, and appears to be a worthwhile preliminary to a full time-dependent model of the system including a realistic description of feedback.

Beside the main cavity, we also assess the role of the short range wake field from geometric aberrations in the vacuum chamber, and
higher order modes in the HHC. These effects are added with the help of improvements in notation and organization of the equations.

Extensive numerical results are reported with parameters for ALS-U. For consistency with the previous work the parameters chosen are partially out of date as the machine design stands at present, but that will not greatly affect our pricipal conclusions.  Although the qualitative picture of the model with HHC alone is still in place, there are large quantitative changes. Even then, we underestimate the full effects, because we can only get convergence
of our iterative method when the current is a few percent less than the design current.

In Section \ref{section:method} we briefly recall our previous algorithm for solving the coupled Ha\" issinski equations. Section
\ref{section:multiple} introduces the improved notation and organization which allows an easy inclusion of multiple resonator wake fields.
Section \ref{section:fullsys} enlarges the system of equations to provide a calculation of the diagonal terms in the potential, thus
overcoming a limitation of the previous formulation. Section \ref{section:adjust} describes the method for determining the generator parameters so as to compensate the induced voltage in the main cavity. Section \ref{section:numerical}, with several subsections, reports numerical results
for the case of ALS-U \cite{alsu1,alsu2}, always making comparisons to results with only the HHC in place. Subsection \ref{subsection:complete} treats the case
of a complete fill, illustrating the compensation of the main cavity in the simplest instance. Subsection \ref{subsection:partial} considers
a partial fill with distributed gaps, as proposed for the machine. Subsection \ref{subsection:smalldf} is concerned with over-stretching by
reduction of the HHC detuning. Subsection \ref{subsection:short} explores the effect of the short range wake field with a realistic
wake potential. Subsection \ref{subsection:hom} checks the effect of the principal higher order mode of the HHC.  Subsection \ref{subsection:full} presents our closest approach to a realistic model, including the harmonic cavity, the compensated main cavity, and the short range wake, altogether. Subsection \ref{subsection:single} examines the effect of the main cavity when there is only a single gap in the bunch train.
 Section \ref{section:outlook} reviews our conclusions and possibilities for further work. Appendix \ref{section:appA} derives the expression for the diagonal terms in the potential, for resonators of arbitrary $Q$.
\section{Summary of method to compute the equilibrium charge densities\label{section:method}}
In \cite{prabI} we derived a set of equations to determine the equilibrium charge densities of $n_b$ bunches, which may be stated succinctly as follows:
\be
F(\hat\rho,I)=0\ . \label{eqns}
\ee
Here $I$ is the average current and $\hat\rho$ is a vector with $2n_b$ real components, consisting  of the real and imaginary parts of $\hat\rho_i(k_{r3})$, where $k_{r3}$ is the wave number of the lowest resonant mode of the 3rd harmonic cavity. These quantities are
defined in terms of the beam frame charge densities $\rho_i(z)$, normalized to 1 on its region of support $[-\Sigma,\Sigma]$, as
\be
\hat\rho_i(k)=\frac{1}{2\pi}\int_{-\Sigma}^\Sigma\exp\big(-ikz(1+i/2Q)\big)\rho_i(z)dz\ ,  \label{rhohat}
\ee
where $Q$ is the quality factor of the cavity. The vector in (\ref{eqns}) is arranged as follows:
\be
\hat\rho=\big[~\rep\hat\rho_1(k_{r3}),\cdots,\rep\hat\rho_{n_b}(k_{r3}),~\imp\hat\rho_1(k_{r3}),\cdots, \imp\hat\rho_{n_b}(k_{r3})~\big]\ . \label{vecrho}
\ee
Accordingly, $F$ in (\ref{eqns}) is a real vector with $2n_b$ components, so that we have $2n_b$ nonlinear algebraic equations in $2n_b$ unknowns,
depending on the parameter $I$.

For the high $Q$ of a typical HHC the quantity (\ref{rhohat}) is very close to the Fourier transform, but we have persistently written all equations for general $Q$ for later applications involving low-$Q$ resonators.

In (\ref{eqns}) the diagonal terms of the induced voltage have been dropped, i.e. the effects on a bunch of its own excitation of the cavity.
This omission is justified  for the typical high $Q$ of an HHC. Our method to handle the diagonal terms in the general case is introduced in
Section \ref{section:fullsys}.

A solution $\hat\rho$ of (\ref{eqns}) determines the charge densities by the formula of Eq.(50) in \cite{prabI},
\be
\rho_i(z_i)=\frac{1}{A_i}\exp\big[-\mu U_i(z_i)\big]\ , \label{rho}
\ee
where $U_i$ is the potential felt by the $i$-th bunch, defined in Eq.(51) of \cite{prabI}.  Here $\mu$ and $A_i$ are constants, and $z_i$ is the beam frame longitudinal coordinate of the $i$-th bunch. The potential $U_i$ depends on all components of $\hat\rho$, on the mean energy loss per turn $U_0$, and on the parameters of the applied voltage $V_{rf}$ which we write as
\be
V_{rf}(z)=V_1\sin(k_1z+\phi_1)=V_1\big(\cos\phi_1\sin(k_1z)+\sin\phi_1\cos(k_1z)\big)\ . \label{vrf}
\ee

We solve (\ref{eqns}) by the matrix version of Newton's iteration, defined in (67) of \cite{prabI}. We begin at small current $I$, taking all components of the first guess for $\hat\rho$ to be the  transform  (\ref{rhohat}) of a Gaussian with the natural bunch length. We then continue step-wise to the desired current, making a linear extrapolation in current to provide a starting guess
for the next Newton iteration at incremented  current. The extrapolation is accomplished by solving for $\ptl \hat\rho/\ptl I$ from the
$I$-derivative of (\ref{eqns}):
\be
\frac{\ptl F}{\ptl\hat\rho}\frac{\ptl\hat\rho}{\ptl I}+\frac{\ptl F}{\ptl I}=0\ .     \label{Ideriv}
\ee

\section{Formalism for Multiple Resonators \label{section:multiple}}
The scheme  allows the inclusion of any number of resonator wake fields, but to do that conveniently requires some care in notation
and organization of the equations. With $n_r$ resonators there are $2n_bn_r=n_u$ unknowns, which we assemble in one long vector $\td\rho$~:
\bea
&&\td\rho=\big[~\td\rho(k)\ ,\ k=1,\cdots,n_u~\big]=    \nonumber\\
&&\big[~\rep\hat\rho_1(k_{r1}),\cdots,~\rep\hat\rho_{n_b}(k_{r1}),~\imp\hat\rho_1(k_{r1})~,
\cdots,~\imp\hat\rho_1(k_{r1})~,\cdots,\nonumber\\
&&~~\rep\hat\rho_1(k_{r,n_r}),\cdots,~\rep\hat\rho_{n_b}(k_{r,n_r})~,
~\imp\hat\rho_1(k_{r,n_r}),\cdots,~\imp\hat\rho_{n_b}(k_{r,n_r})~\big]\ .
\label{tilderho}
\eea
Here $k_{r,n}$ is the resonant wave number of the $n$-th resonator, and the subscript of $\hat\rho$ denotes as usual the bunch number.

To identify the bunch number and the resonator number for the $k$-th component of the vector, we define two index maps: $\iota(k)$ which gives the
bunch number and $r(k)$ which gives the resonator number. Namely,
\bea
&&\iota(k)=\left\{
\begin{array}{l}\mod(k,n_b)~~{\rm if}\mod(k,n_b)\ne 0\\
~~~n_b~~{\rm if}\mod(k,n_b)= 0\\ \end{array}\right\} \label{io}\\
 &&r(k)=\bigg\lceil \frac{k}{2n_b}\bigg\rceil\ . \label{rk}
\eea
Here $\lceil x\rceil$, the ceiling of $x$,  is the least integer greater than or
equal to $x$. We also need two projection operators: $P_{re}(k)$ which is equal to 1 if $k$ corresponds to a $\rep\hat\rho$ and is zero otherwise, and
$P_{im}(k)$ which is equal to 1 if $k$ corresponds to a $\imp\hat\rho$ and is zero otherwise. These are expressed in terms
of the ceiling of $k/n_b$ as follows:
\bea
&&P_{re}(k)=\frac{1}{2}\bigg[1-(-1)^{\lceil k/n_b \rceil}\bigg]\ ,\nonumber\\
&&P_{im}(k)=\frac{1}{2}\bigg[1+(-1)^{\lceil k/n_b \rceil}\bigg]\ . \label{projectors}
\eea

The potential $U_j(z)$ for bunch $j$, generalizing Eq.(51) of \cite{prabI} to allow $n_r$ resonators,
is stated as
\bea
&&U_j(z)=\frac{eV_1}{k_1}\big[x_1\cos(k_1z)-x_2\sin(k_1z)-x_1\big]+U_0z\label{udef1}\\
&&+\sum_{n=1}^{n_r} U^d_{jn}(z)+\sum_{k=1}^{n_u}M(z)_{j,k}~\td\rho_k\ ,\quad
j=1,\cdots,n_b\ ,\quad -\Sigma~\le~ z~\le~\Sigma\ .\label{udef2}
\eea
The first term in (\ref{udef1}) is $e$ times the integral of the applied voltage, now called the generator voltage and written as
\be
V_g(z)=V_1\big[x_1\sin(k_1z)+x_2\cos(k_1z)\big]\ .  \label{vgen}
\ee
At $x_1=\cos\phi_1,\ x_2=\sin\phi_1$ this reduces to the desired $V_{rf}$ of (\ref{vrf}). In an amplitude-phase representation
we have
\be
V_g(z)=\td V_1\sin(k_1z+\td\phi_1)\ ,\quad \td V_1=(x_1^2+x_2^2)^{1/2}V_1\ ,\quad \td\phi_1=\tan^{-1}(x_2/x_1)\ . \label{phamp}
\ee
The first term in (\ref{udef2}) represents the diagonal contributions, the effect on bunch $j$ of its own excitation of the resonators, as opposed to excitation by the other bunches which is described by the second term. By writing the latter as a simple
matrix-vector product we greatly simplify the calculation of the Jacobian of the system, making it formally the same for any number
of resonators.

Referring to Eqs.(28), (51), (55), (56), (57), (58) of \cite{prabI}, we can write down the matrix elements $M(z)_{i,k}$ in
the second term of (\ref{udef2}). For this we introduce  a notation appropriate for labeling by the index $k$ of (\ref{tilderho}). Functions of $k$, defined
via the index maps, are labeled with a tilde:
\bea
&&\tilde k_{r,k}=k_{r,r(k)}\ , \nonumber\\
&&\tilde\xi_k=\xi_{\iota(k)}\ ,\quad \td A_k=A_{\iota(k)} \nonumber\\
&&\tilde R_{sk}=R_{s,r(k)}\ ,\quad
\tilde Q_k=Q_{r(k)}\ ,\nonumber\\
&&\tilde\eta_k=\eta_{r(k)} ,\quad
\tilde\psi_k=\psi_{r(k)}\ ,\nonumber\\
&&\tilde \phi_{j,k}=\tilde k_{r,k}\big[ (m_{\iota(k)}-m_j)\lambda_1+\theta_{j-1,\iota(k)}C\big]\ ,\nonumber\\
&&\sigma_{j,k}(z)=\mathcal{S}\big(\tilde k_{r,k}z,\ \tilde Q_k,\ \tilde\phi_{j,k}+\tilde\psi_k\big)\ ,\nonumber\\
&&\gamma_{j,k}(z)=\mathcal{C}\big(\tilde k_{r,k}z,\ \tilde Q_k,\ \tilde\phi_{j,k}+\tilde\psi_k\big)\ ,\label{compact}
\eea
where
\bea
&&\mathcal{S}(k_rz,Q,\phi)=\frac{1}{1+(1/2Q)^2}\bigg[\exp(-k_rz/2Q)\bigg(\sin(k_rz+\phi)-\frac{1}{2Q}\cos(k_rz+\phi)\bigg)\bigg]_0^z\ ,\nonumber\\
&&\mathcal{C}(k_rz,Q,\phi)=\frac{1}{1+(1/2Q)^2}\bigg[\exp(-k_rz/2Q)\bigg(\cos(k_rz+\phi)+\frac{1}{2Q}\sin(k_rz+\phi)\bigg)\bigg]_0^z\ .\nonumber\\
\label{calSC}
\eea
The result for the matrix from (51) and (57) of \cite{prabI} is seen to be (noting that $\omega_r/k_r=c$)
\be
M(z)_{j,k}=2\pi ce^2N \frac{\td\eta_j\tilde R_{sj}}{\td Q_j}(1-\delta_{j,\iota(k)})\td\xi_k
\exp(-\td\phi_{j,k}/2\td Q_k)\big[
P_{re}(k)\sigma_{j,k}(z)+P_{im}(k)\gamma_{j,k}(z)\big] \ .\label{mdef}
\ee

In the present notation the system of coupled Ha\"issinski equations, generalizing (66) of \cite{prabI}, takes the form
\bea
&&F_j(\td\rho)=\td A_j\td\rho_j-\frac{1}{2\pi}\int_{-\Sigma}^\Sigma
\big[P_{re}(k)\cos(\td k_{r,j}\zeta)-P_{im}(k)\sin(\tilde k_{r,j}\zeta)\big]\nonumber\\
&&\hskip 2cm \cdot\exp\big[\td k_{r,j}\zeta/2\td Q_j-\mu\ U_{\iota(j)}(\zeta)\big]d\zeta =0\ ,\quad j=1,\cdots,n_u\ .\label{hais}
\eea
The normalization integral appearing in the first term is
\be
\td A_j= \int_{-\Sigma}^\Sigma\exp\big[ -\mu\ U_{\iota(j)}(\zeta)\big]d\zeta\ . \label{Adef}
\ee

We require the Jacobian matrix $[\ptl F_j/\ptl \tilde\rho_k]$ for the solution of (\ref{hais}) by Newton's method,
assuming that the diagonal terms are fixed. This is found immediately from (\ref{udef2}), (\ref{hais}), and
(\ref{Adef}) as
\bea
&&\frac{\ptl F_j}{\ptl\tilde\rho_k}=\td A_j\delta_{j,k}-\mu\int_{-\Sigma}^\Sigma\exp\big[ -\mu U_{\iota(j)}(\zeta)\big]M(\zeta)_{j,k}\nonumber\\
&&\cdot\bigg[\td\rho_j-\frac{1}{2\pi}\big[P_{re}(k)\cos(\tilde k_{r,j}\zeta)-P_{im}(k)\sin(\tilde k_{r,j}\zeta)\big]
\exp\big[\td k_{r,j}\zeta/2\td Q_j\big]\bigg]d\zeta\ .\label{jacobian}
\eea

The compact expressions in (\ref{mdef}), (\ref{hais}), and (\ref{jacobian}) are quite convenient for coding, and
lead to a short program to solve the Ha\"issinski equations with any number of resonators. For $\zeta$ at $n_p$ mesh points $z_i$ used in the integrals we have the  array $M(i,j,k)=M(z_i)_{j,k}$ of manageable dimension $n_p\times n_b\times n_u$ which can be computed and stored at the top, outside the Newton iteration.

For the work of the following section we also need the induced voltage from the main cavity, which we designate as the first resonator
in the list $(n=1)$. For the $j$-th bunch this takes the form
\bea
&&V_{r1j}(z)=-2\pi ceN\frac{k_{r1}R_{s1}\eta_1}{Q_1}\bigg[\sum_{k=1}^{2n_b}(1-\delta_{j,\iota(k)})\td\xi_k
\exp(-(\td k_{r,k}z+\td\phi_{j,k})/2\td Q_k)\nonumber\\&&
\bigg(P_{re}(k)\cos(\td k_{r,k}z+\td\phi_{j,k}+\td \psi_k)-P_{im}(k)\sin(\td k_{r,k}z+\td\phi_{i,k}+\td \psi_k)\bigg)\td\rho_k
+v^d_{1j}(z)\bigg]\ .\nonumber\\ \label{vr2}
\eea
The diagonal term $v^d_{1j}$ can be evaluated in terms of integrals derived in Appendix \ref{section:appA}.
\section{The full system of equations with diagonal terms \label{section:fullsys}}
Through (\ref{hais}) we have a system of $n_u$ algebraic equations for determination of $\td\rho$, provided that the diagonal terms in $U_i$ are given.
The latter are functionals of the charge densities $\rho_i(z_i)$, from which it follows that (\ref{hais}) can be stated in vector notation as
\be
\td\rho=\mathcal{A}(\td\rho,\rho,I)\ .\label{Aeqn}
\ee
On the other hand, the $\rho_i(z_i)$ are determined in turn
as solutions of integral equations provided that $\td\rho$ is given.  The integral equations are like normal single-bunch Ha\" issinski equations, but with a background potential determined by $\td\rho$, namely
\be
\rho_i(z_i)=\frac{1}{A_i}\exp\bigg[-\mu U_i(z_i,~\rho_i,~\td\rho)\bigg]\ ,\quad i=1,\cdots,n_b\ .   \label{hais1}
\ee
In vector notation
\be
\rho=\mathcal{B}(\rho,\td\rho,I)\ . \label{Beqn}.
\ee
The potential $U_i$ depends on the $\rho_i$ through its diagonal terms, in the first sum in (\ref{udef2}).
Our procedure will be to interleave the solution of (\ref{Aeqn}) at fixed $\td\rho$, by the usual Newton method,  with the  solution of (\ref{Beqn}) at fixed $\td\rho$ . If this algorithm converges we shall have consistency between $\rho$ and $\td\rho$ and a solution of the full system.

It turns out, most fortunately, that the solution of (\ref{Beqn}) is obtained by plain iteration as would be applied to a contraction mapping,
\be
\rho^{(n+1)}=\mathcal{B}(\rho^{(n)},\td\rho,I)\ .     \label{plain}
\ee
In our application this usually converges to adequate accuracy in just one step, or three at most, and takes negligible time.

This scheme based on (\ref{Aeqn}) and (\ref{Beqn}) is used in all calculations reported below. It replaces the method used in \cite{prabI}, which
was to evaluate the diagonal terms from the value of $\rho$ from the previous Newton iterate. That works only for high-$Q$ resonators,
so is not adequate for handling the short range machine wake.

\section{Algorithm to adjust the generator parameters $(x_1,x_2)$ \label{section:adjust}}

We wish to choose $(x_1,x_2)$ so as to minimize, in some sense, the difference
\be
V_{rf}(z_i)-V_g(z_i,x_1,x_2)-V_{r1i}(z_i,x_1,x_2)\ ,\label{tomin}
\ee
for all $i=1,\cdots,n_b$.  A reasonable and convenient choice for an objective function to minimize is the sum of the squared $L^2$ norms of the quantities
(\ref{tomin}).  With a normalizing factor to make it dimensionless and of convenient magnitude that is
\bea
&&f(x_1,x_2)=\nonumber\\
&&\frac{1}{2\Sigma V_1^2}\sum_{i=1}^{n_b} \int_{-\Sigma}^\Sigma \bigg[V_1(\cos\phi_1-x_1)\sin(k_1z)+ V_1(\sin\phi_1-x_2)\cos(k_1z)
-V_{r1i}(z,x_1,x_2)\bigg]^2dz\ . \nonumber\\ \label{obj}
\eea
The region of integration $[-\Sigma,\Sigma]$ is the same as that used in the definition of the potential $U_i$.

Note that the minimum of $f$ cannot be strictly zero, since $V_{r1}$ is sinusoidal with wave number $k_{r1}$, whereas the other
terms are sinusoidal with a slightly different wave number $k_1$.

Let us adopt the vector notation $x=(x_1,x_2)$ with norm $|x|=|x_1|+|x_2|$.
The equations to solve now depend on $x$, having the form
\be
F(\td\rho,I,x)=0 \ .  \label{eqnsx}
\ee
To avoid notational clutter we suppress reference to the diagonal terms, leaving it understood that a solution of (\ref{eqnsx}) for $\td\rho$ actually involves the scheme of the previous session.
As usual we solve for $\td\rho$, for an increasing sequence of $I$-values. {\it The scheme will be to
minimize $f(x)$ at each $I$, thus providing a new $x=\arg\min f$ to be used at the next value of $I$.} As will now be explained, the
minimization will also be done iteratively, so that we have an $x$-iteration embedded in the $\td\rho$-iteration.

We wish to zero $\nabla_x F$, which is to find $x$ to solve the equations
\bea
&& \sum_{i=1}^{n_b} \int_{-\Sigma}^\Sigma
 \bigg[V_1(\cos\phi_1-x_1)\sin(k_1z)+ V_1(\sin\phi_1-x_2)\cos(k_1z)
-V_{r1i}(z,x)\bigg]\nonumber\\
 &&\times \left[\begin{array} {c} V_1\sin(k_1z)+\ptl_{x_1} V_{r1i}(z,x) \\
V_1\cos(k_1z)+\ptl_{x_2} V_{r1i}(z,x)\end{array}\right]dz=
\left[\begin{array}{c}0\\0\end{array}\right]\ .
\nonumber\\ \label{grad0}
\eea

To solve (\ref{grad0}) a first thought might be to apply Newton's method, starting at some low current and choosing the
zero current solution $(\cos\phi_1,\sin\phi_1)$ as the first guess. This would be awkward, however, since it would involve the
second derivatives of $V_{r1i}$ with respect to $(x_1,x_2)$.  The first derivatives must already be done by an expensive numerical differentiation, and the second numerical derivative would be error prone and even more expensive. Instead, let us assume that we have a first guess
$(x_{10},x_{20})$ and suppose that in a small neighborhood of that point the first derivatives of $V_{r1i}$ can be regarded as constant.
Then second derivatives are zero and the Taylor expansion of $V_{r1i}$ gives two linear equations to solve for $(x_1,x_2)$, namely
\be
\left[\begin{array}{cc}a_{11}&a_{12}\\a_{21}&a_{22}\end{array}\right]\left[\begin{array}{c} x_1\\x_2\end{array}\right]=
\left[\begin{array} {c} b_1\\b_2\end{array}\right]\ , \label{axb}
\ee
where
\bea
&&a_{11}=\sum_i \int\alpha_{1i}(z,x_0)^2dz\ ,\quad a_{22}=\sum_i \int\alpha_{2i}(z,x_0)^2dz\ ,\nonumber\\
&&a_{12}=a_{21}=\sum_i \int\alpha_{1i}(z,x_0)\alpha_{2i}(z,x_0)dz\ ,\nonumber\\
&&b_1=\sum_i \int\alpha_{1i}(z,x_0)\beta_i(z,x_0)dz\ ,\quad b_2=\sum_i \int\alpha_{2i}(z,x_0)\beta_i(z,x_0)dz\ ,
\nonumber\\ \label{abdef}
\eea
with
\bea
&&\alpha_{1i}(z,x_0)=V_1\sin(k_1z)+\ptl_{x_1}V_{r1i}(z,x_0)\ ,\nonumber\\ &&\alpha_{2i}(z,x_0)=V_1\cos(k_1z)+\ptl_{x_2}V_{r1i}(z,x_0)\ ,\nonumber\\
&&\beta_i(z,x_0)=-V_{r1i}(z,x_0)+\nabla_x V_{r1i}(z,x_0)\cdot x_0+V_1\sin(k_1z+\phi_1)
\\ \label{alfbet}
\eea

By (\ref{axb}) we have an update $x_0\rightarrow x$  which establishes the pattern of the general iterate $x^{(k)}\rightarrow x^{(k+1)}$. This will be carried
to convergence in the sense $|x^{(k+1)}-x^{(k)}|<\epsilon_x$, with a suitable $\epsilon_x$ to be determined by experiment. Each iterate requires
a value for $V_{r1i}$ and for $\nabla_xV_{r1i}$, which we compute numerically by a divided difference,
\be
 \frac{\ptl V_{r1i}}{\ptl x_1}(z,x)\approx \frac{V_{r1i}(z,x_1+\Delta x,x_2)-V_{r1i}(z,x_1,x_2)}{\Delta x}\ . \label{fd}
\ee
Thus one $x$-iteration requires three $\td\rho$-iterations  to provide the necessary values of $V_{r1i}$  (which are constructed
from $\td\rho$). The first $\td\rho$ iteration to find $V_{r1i}(z,x_1,x_2)$ produces a $\td\rho$ which is a very good guess to start the
remaining two iterations to make the derivatives, which then converge quickly.

The choice of $\Delta x$ in (\ref{fd}) requires a compromise between accuracy and avoiding round-off error. We found that $\Delta x=10^{-4}$ was
widely satisfactory, whereas success with smaller values depended on the circumstances.

\section{Numerical results with and without the main cavity \label{section:numerical}}

As in \cite{prabI} we illustrate with parameters for ALS-U \cite{alsu1,alsu2}, the forthcoming Advanced Light Source Upgrade. Although the machine design is not yet final, one provisional
set of parameters for our main cavity (actually the effect of two cavities together) is as follows:
\be
R_s=0.8259~ M\Omega\ ,\quad Q=3486\ ,\quad \delta f=f_{r1}-f_1= -82.54 ~kHz  \label{MCpara}
\ee
Here the shunt impedance $R_s$ and quality factor $Q$ are loaded values, the unloaded values divided by $1+\beta$, with
coupling parameter $\beta=7.233$. We take these parameters for the main cavity, otherwise keeping the same parameters as in
\cite{prabI}, Table I.  Thus we take $U_0=$217 keV, even though a value of 330 keV may  be contemplated for the set (\ref{MCpara}).
\subsection{Complete Fill \label{subsection:complete}}
We first take the case of a complete fill, thus $n_b=h=328$.
The average current is to be 500 mA, which we reach in 8 steps starting from 200 mA. The CPU time is 15 minutes, rather than 20 seconds
for the calculation without the main cavity. The increase is mostly due to a much slower convergence of the $\td\rho$-iteration,
the $x$-iteration being a minor factor in CPU time. To save time we gave $\epsilon_x$ the rather large value of 0.05, but then made a refinement to $\epsilon_x=10^{-6}$ at the final current, in an extra 2 minute. The steepness of the objective function $f(x1,x2)$ of (\ref{obj}) is extraordinary, having values around $10^4$ in the sequence with $\epsilon_x=0.05$ while falling to a value close to 1 after the refinement.
An interesting question is how this steepness would be reflected in a feedback system.

\begin{figure}[htb]
   \centering
   \includegraphics[width=.6\linewidth]{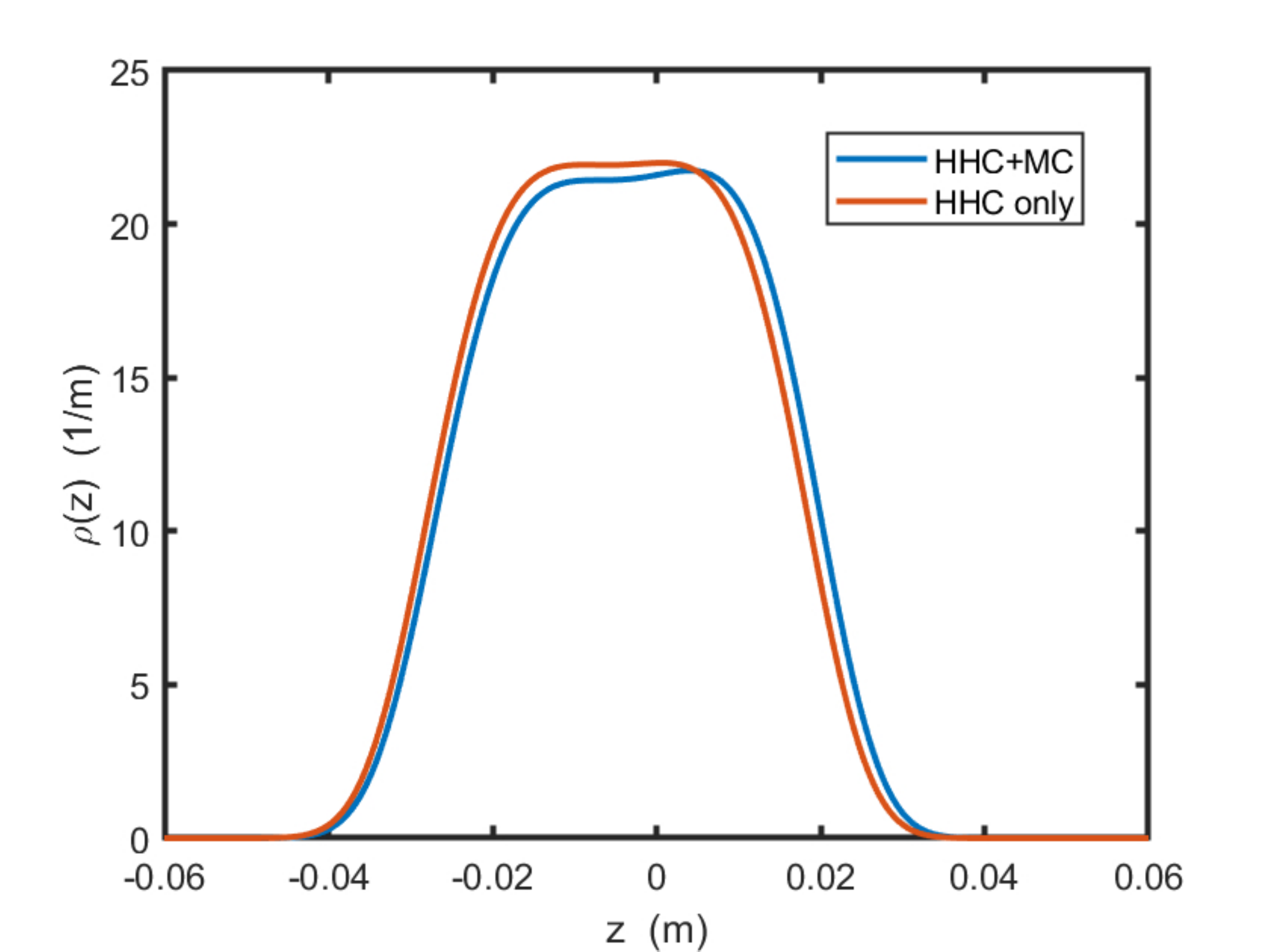}  %rho_328_HHC_MC_I500.pdf
  \caption{Charge density for complete fill at 500 mA, with compensated main cavity (blue)
  and without main cavity (red).}
   \label{fig:fig1}
\end{figure}

The result for the charge density, shown in the blue curve of Fig.\ref{fig:fig1}, is quite close to the result without the main cavity,
shown in red. It should be emphasized that there is no explicit constraint requiring all bunches to be the same. We have computed 328 bunches
separately, and have found that they all come out to be the same. This constitutes a good check on the correctness of the equations and the code.

\begin{figure}[htb]
   \centering
   \includegraphics[width=.6\linewidth]{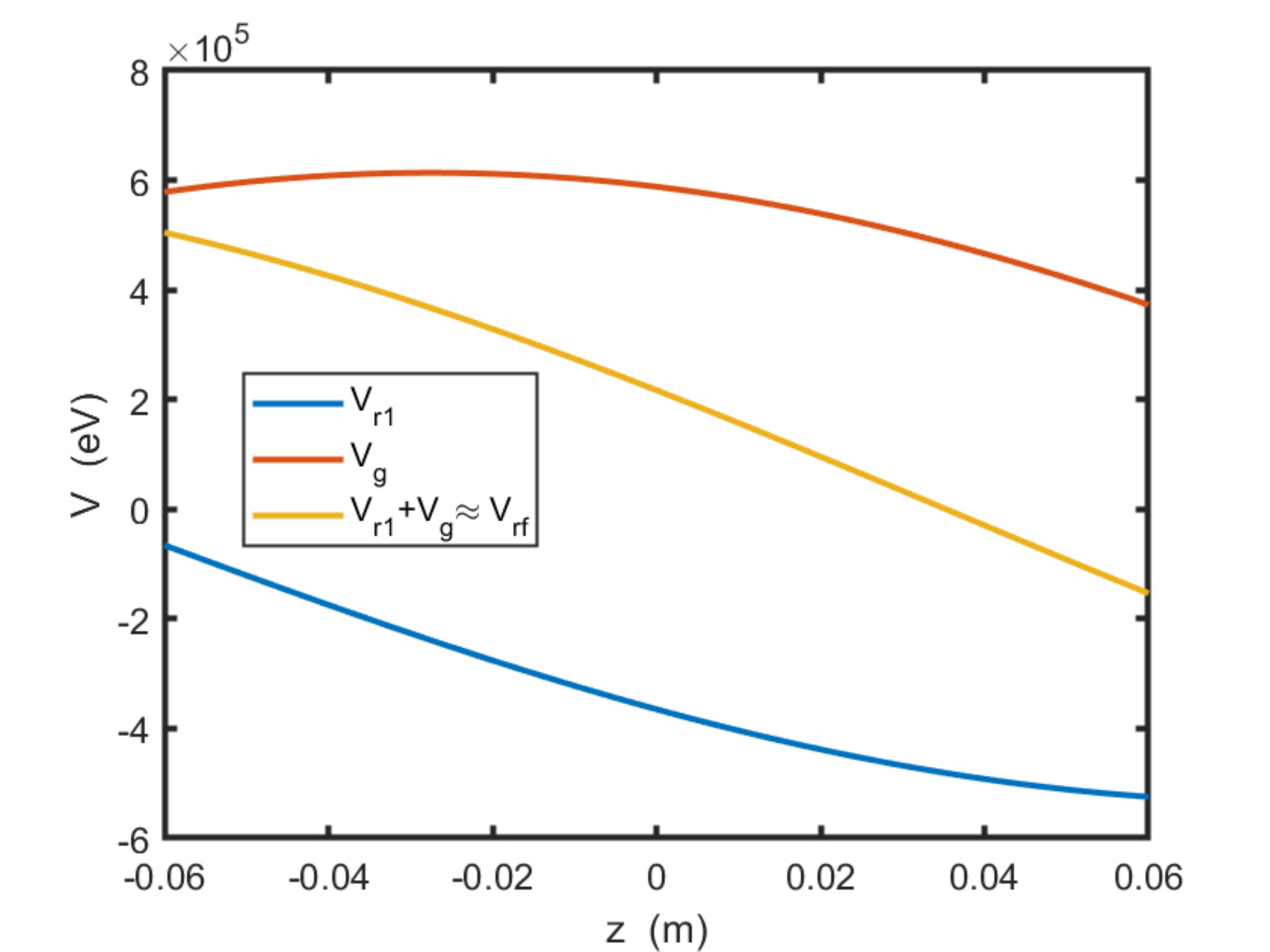}    %vmc_328_HHC_MC.pdf
  \caption{MC induced voltage $V_{r1}$, generator voltage $V_g$ and their sum.}
   \label{fig:fig2}
   \end{figure}

\begin{figure}[htb]
    \includegraphics[width=.6\linewidth]{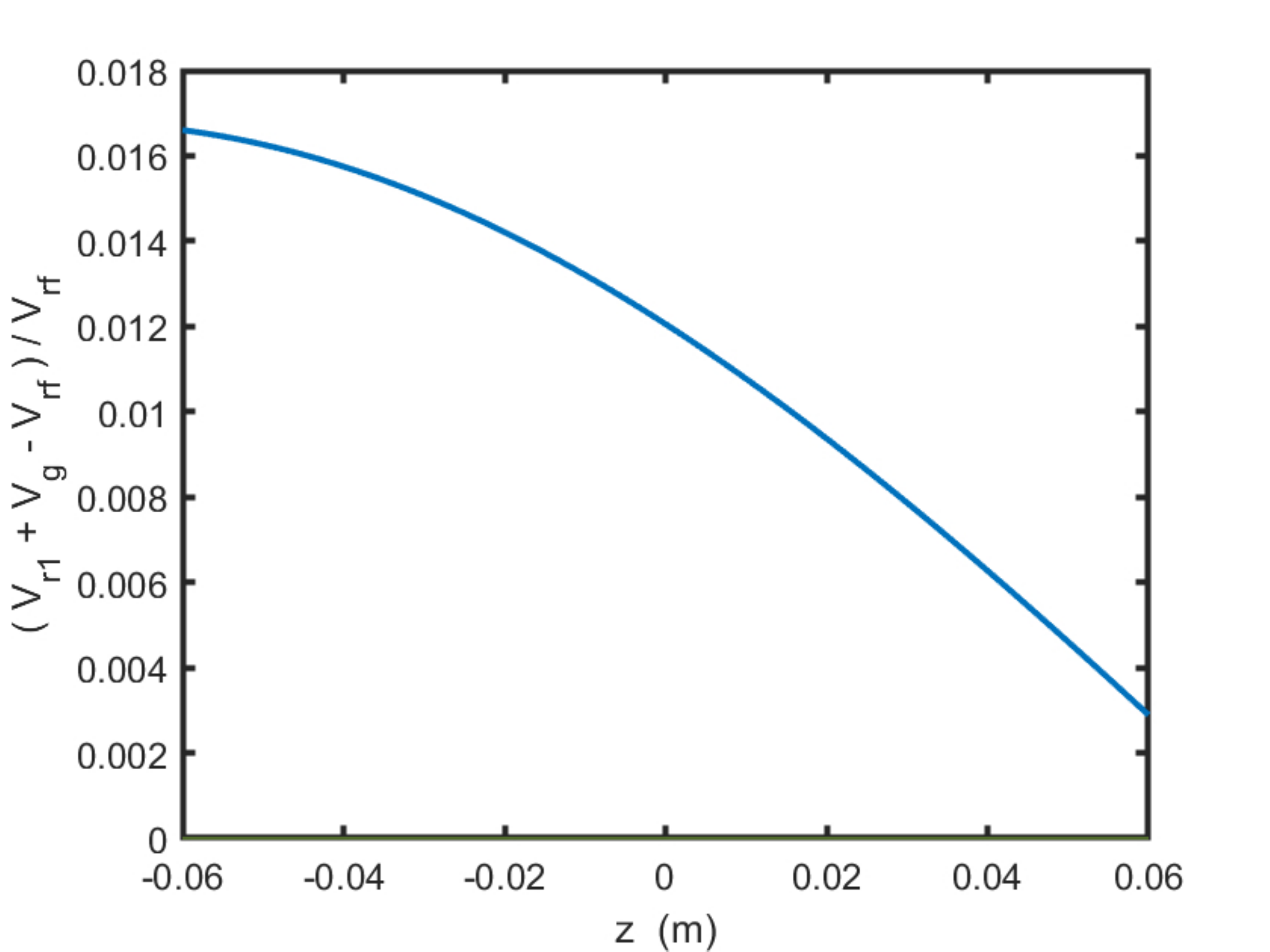}    %dv_328_HHC_MC_I500.pdf
   \caption{Relative deviation of $V_{r1}+V_g$ from $V_{rf}$.}
   \label{fig:fig3}
\end{figure}

In Fig.\ref{fig:fig2} we show the compensation mechanism. The sum of the generator voltage $V_g$ and the induced voltage $V_{r1}$ from the main cavity  is the orange curve. The latter deviates from the desired effective voltage $V_{rf}$ by less than 2\%, as is seen
Fig.\ref{fig:fig3}.

The phasor of the generator voltage moves closer to $\pi/2$ and its magnitude $\sqrt{x_1^2+x_2^2}$ increases from 1 to 1.0245, in comparison to the phasor of $V_{rf}$. \break The corresponding values of $(x_1,x_2)$ are
\be
 (x_{10},x_{20})=(\cos\phi_1,\sin\phi_1)= (~-0.93231,~0.36167~)   \quad\rightarrow\quad (x_1,x_2)=(~-0.29098,~0.98018~)\ .
\ee
\subsection{Partial fill C2 with distributed gaps \label{subsection:partial}}
Next we take a partial fill with distributed gaps, labeled as fill C2; see Section XIII-C of \cite{prabI}. There are 284 bunches in 11 trains, with 4 empty buckets between trains. There are 9 trains of 26 and 2 of 25, with the latter positioned at opposite sides of the ring. All bunches have the same charge. As in the preceding example we start the calculation at low average current and advance in steps trying to reach the desired 500 mA. The convergence of iterations is at first  similar to that of the preceding case, but begins to falter around 430 mA average current, at which point the convergence of the $\td\rho$-iteration becomes problematic. By taking smaller and smaller steps in current we can reach 496 mA, but beyond that point the Jacobian matrix of the system appears to approach a singularity, as is indicated by its estimated condition number having a precipitous increase, from 700 at the last good solution to 2900 at a slightly higher current.  Nevertheless, the $x$-iterations continue to converge as long as the $\td\rho$-iterations do. In the following, graphs are plotted for the maximum achievable current, stated in figure captions.

{\bf Now the plots of $V_g$ and $V_{r1}$  and their sum look exactly the same as in Fig.\ref{fig:fig2}, for every bunch.
The minimization of $f(x_1,x_2)$ has caused the bunch forms to rearrange themselves so that the compensation is essentially perfect
for every bunch. The deviation of $V_{r1}+V_g$ from $V_{rf}$, scarcely visible on the scale of Fig.\ref{fig:fig2}, varies from bunch to bunch, but is still less than 3\% for all bunches.}

\begin{figure}[htb]
   \centering
   \begin{minipage} [b]{.49\linewidth}
   \includegraphics[width=\linewidth]{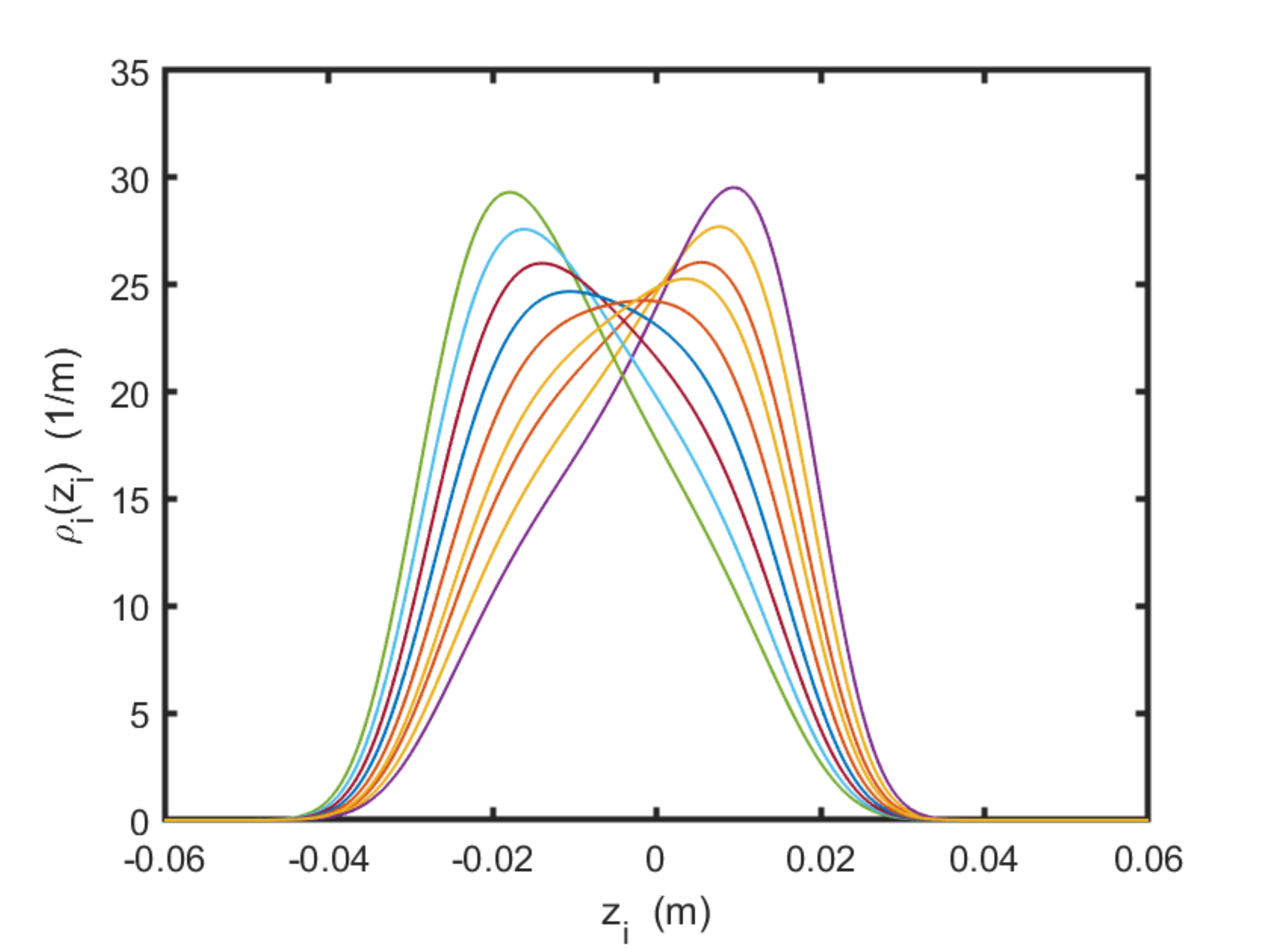}    %rho_284_ng_HHC_MC_I0573.pdf
  \caption{Charge densities in a train of 26,\\ surrounded by gaps of 4 buckets, fill C2,\\ MC beam loading included, $I_{\rm av}=496$ mA.}
   \label{fig:fig4}
   \end{minipage}
   \begin{minipage} [b]{.49\linewidth}
   \includegraphics[width=\linewidth]{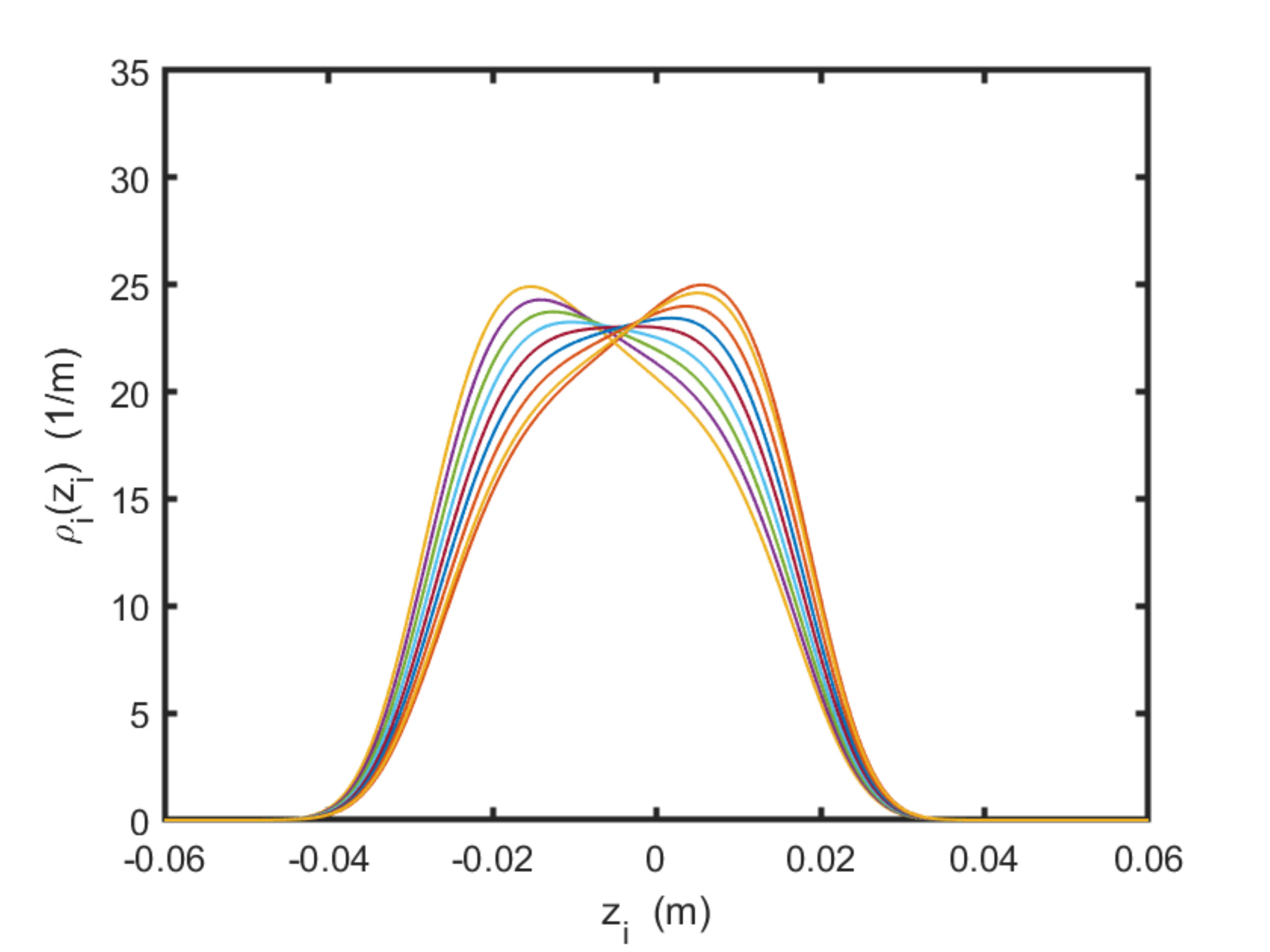}     %rho_284_ng_HHC_I0573.pdf
   \caption{Charge densities in a train of 26,\\ surrounded by gaps of 4 buckets, fill C2,\\ MC beam loading omitted,  $I_{\rm av}=496$ mA.}
   \label{fig:fig5}
   \end{minipage}
\end{figure}

\begin{figure}[htb]
   \centering
   \includegraphics[width=.6\linewidth]{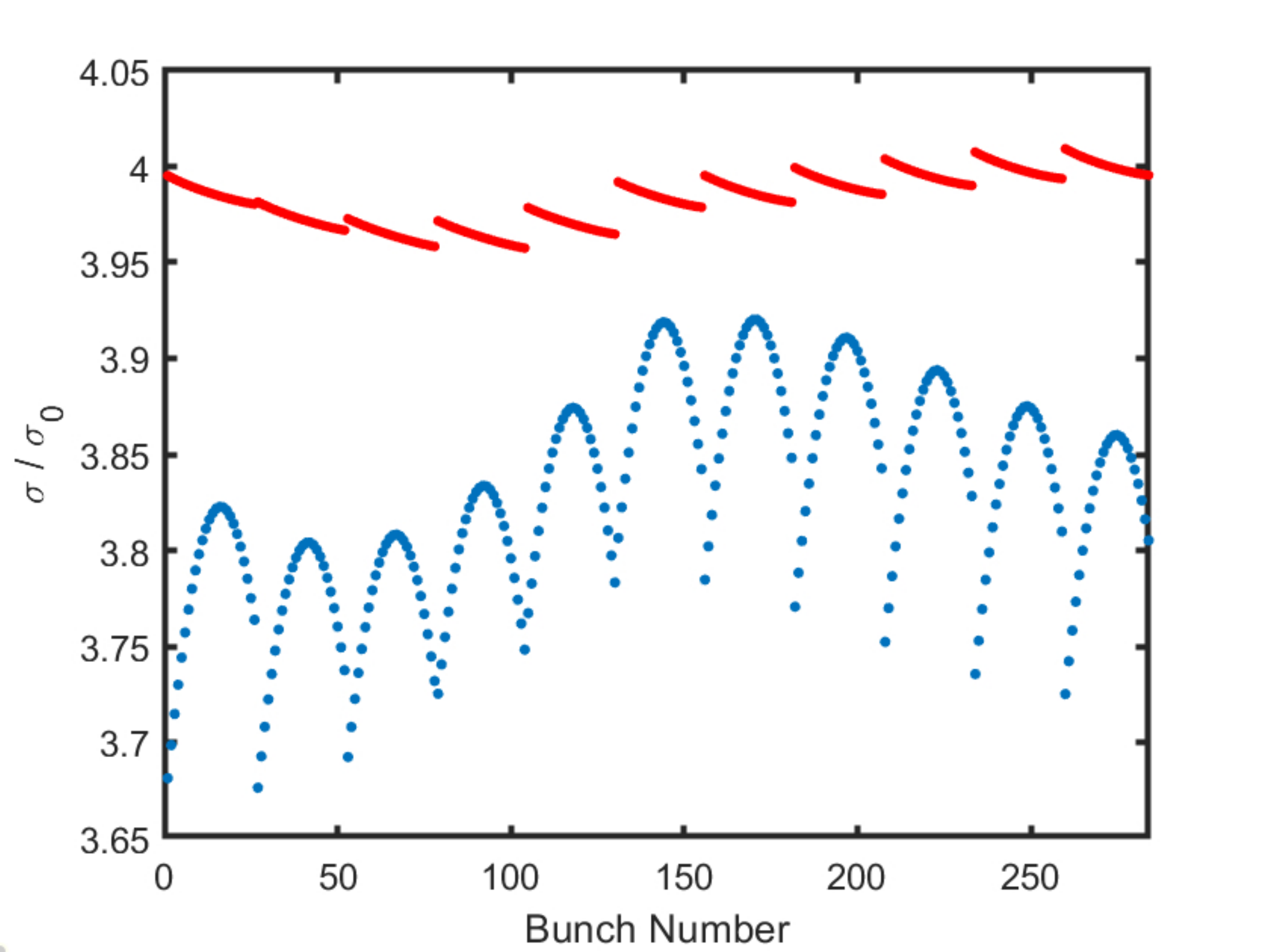}     %sig_284_ng_HHC_MC_I0573.pdf
  \caption{Bunch length increases in a train of 26, surrounded by gaps of 4 buckets,  with main cavity beam loading (blue) and without (red). $I_{\rm av}=496$ mA. The plot is the ratio of bunch length $\sigma$ to the natural bunch length $\sigma_0$.}
   \label{fig:fig6}
\end{figure}

Fig.\ref{fig:fig4} shows 9 bunch profiles in one train, to be compared with the corresponding results without the main cavity in Fig. \ref{fig:fig5}. The main cavity causes considerably more bunch distortion along the train, and also a bigger variation in the rms bunch lengths, as is seen in Fig.\ref{fig:fig6}. The plots show the ratio of bunch length to the natural bunch length. The head of the train is on the right, with the highest bunch number.

\begin{figure}[htb]
   \centering
   \includegraphics[width=.6\linewidth]{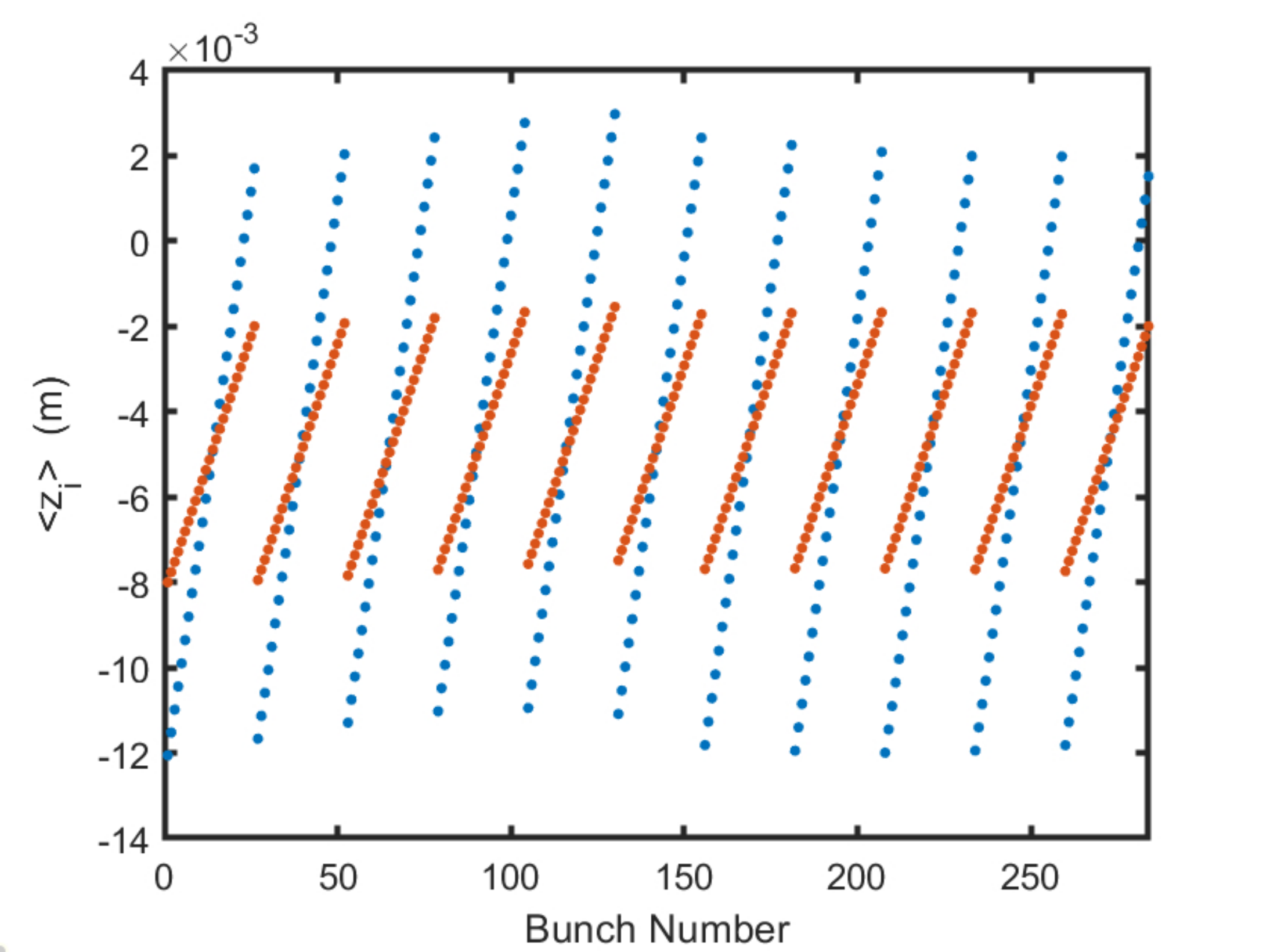}     %zb_284_ng_HHC_MC_I0573.pdf
  \caption{Centroids $<z>$ in a train of 26, surrounded by gaps of 4 buckets,  with cavity beam loading (blue) and without (red). $I_{\rm av}=496$ mA}
   \label{fig:fig7}
\end{figure}

The corresponding results for the bunch centroids is seen in Figs.\ref{fig:fig7}. Again the deviation from the
 case without the main cavity is quite substantial.

\begin{figure}[htb]
   \centering
   \includegraphics[width=.6\linewidth]{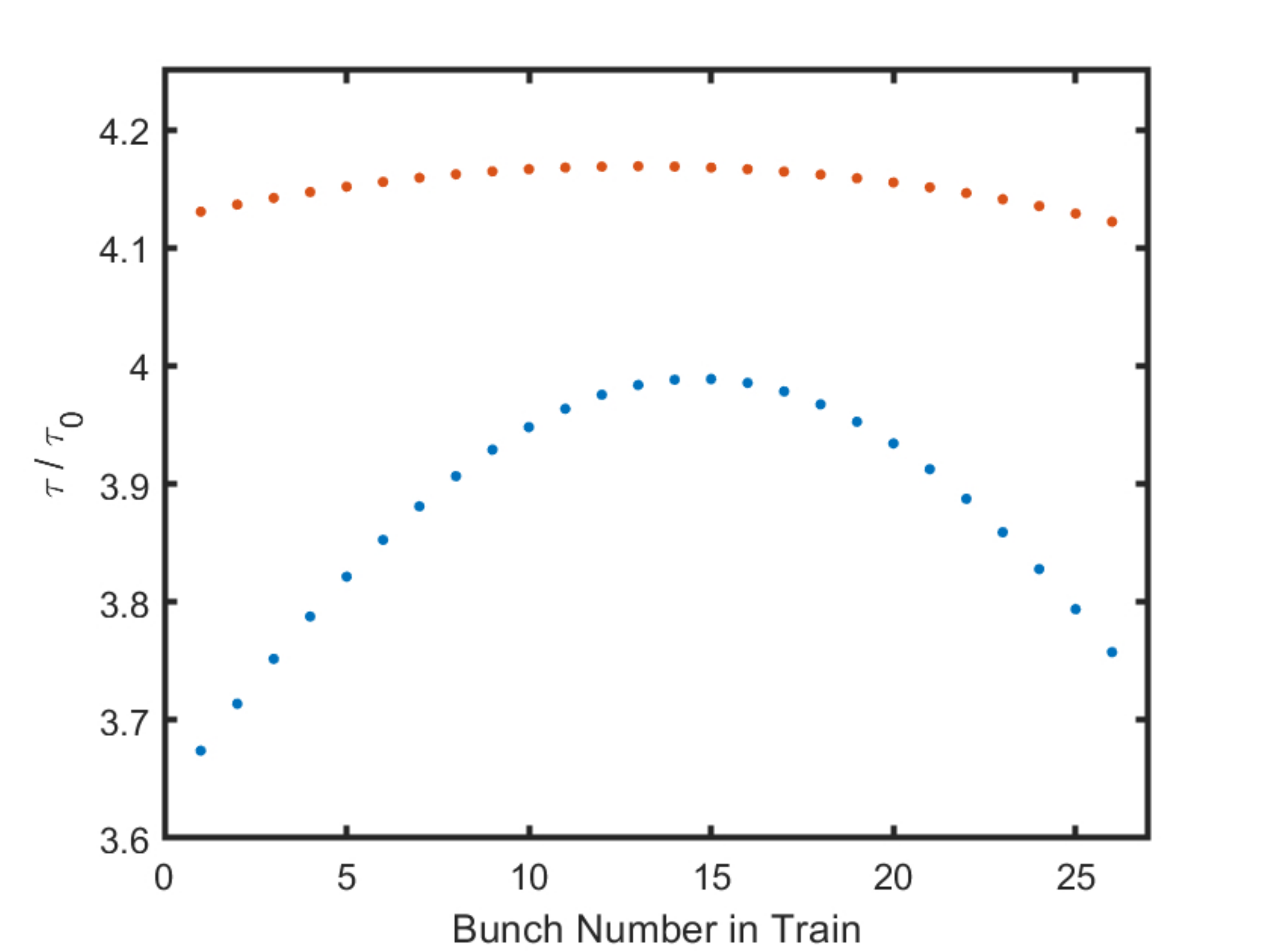}     %tou_284_ng_HHC_MC_I0573.pdf
  \caption{Touschek lifetime increase along a train, with compensated MC (blue) and without (red). $I_{\rm av}=496$ mA.}
   \label{fig:fig8}
\end{figure}

The main point of practical interest is the increase in Touschek lifetime achieved through the bunch stretching caused by the HHC.
Again, the MC has a sizeable effect in reducing the lifetime and in causing a larger variation along a train. This is shown in Fig.\ref{fig:fig8} which
gives the ratio of the lifetime $\tau$ to the lifetime $\tau_0$ without the MC.

\begin{figure}[htb]
   \centering
   \begin{minipage} [b]{.49\linewidth}
   \includegraphics[width=\linewidth]{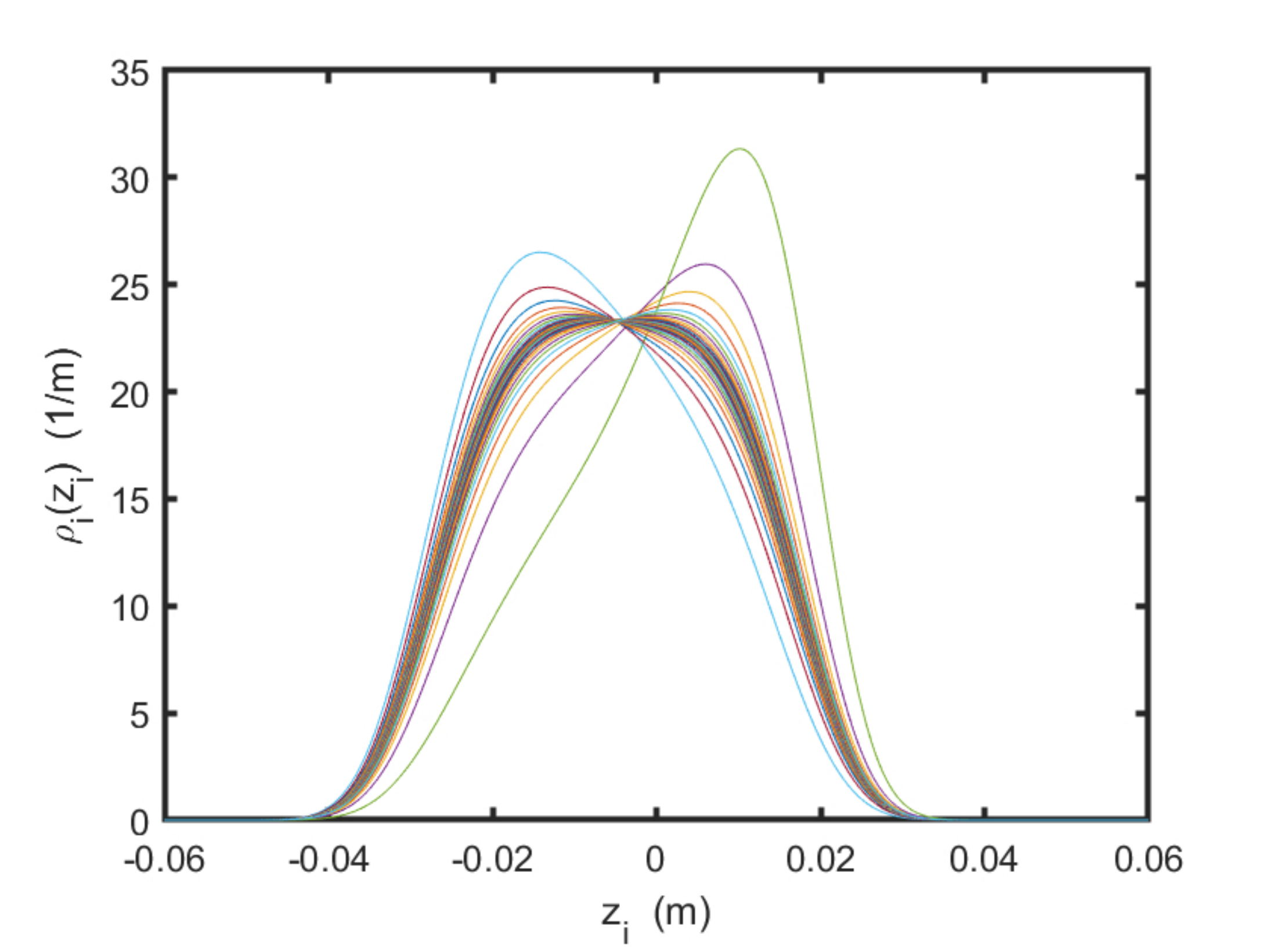}     %rho_284_taper_HHC_MC_I496.pdf
  \caption{Case of tapered bunch charges,\\ MC beam loading included, $I_{\rm av}=496$ mA.}
   \label{fig:fig9}
   \end{minipage}
   \begin{minipage} [b]{.49\linewidth}
   \includegraphics[width=\linewidth]{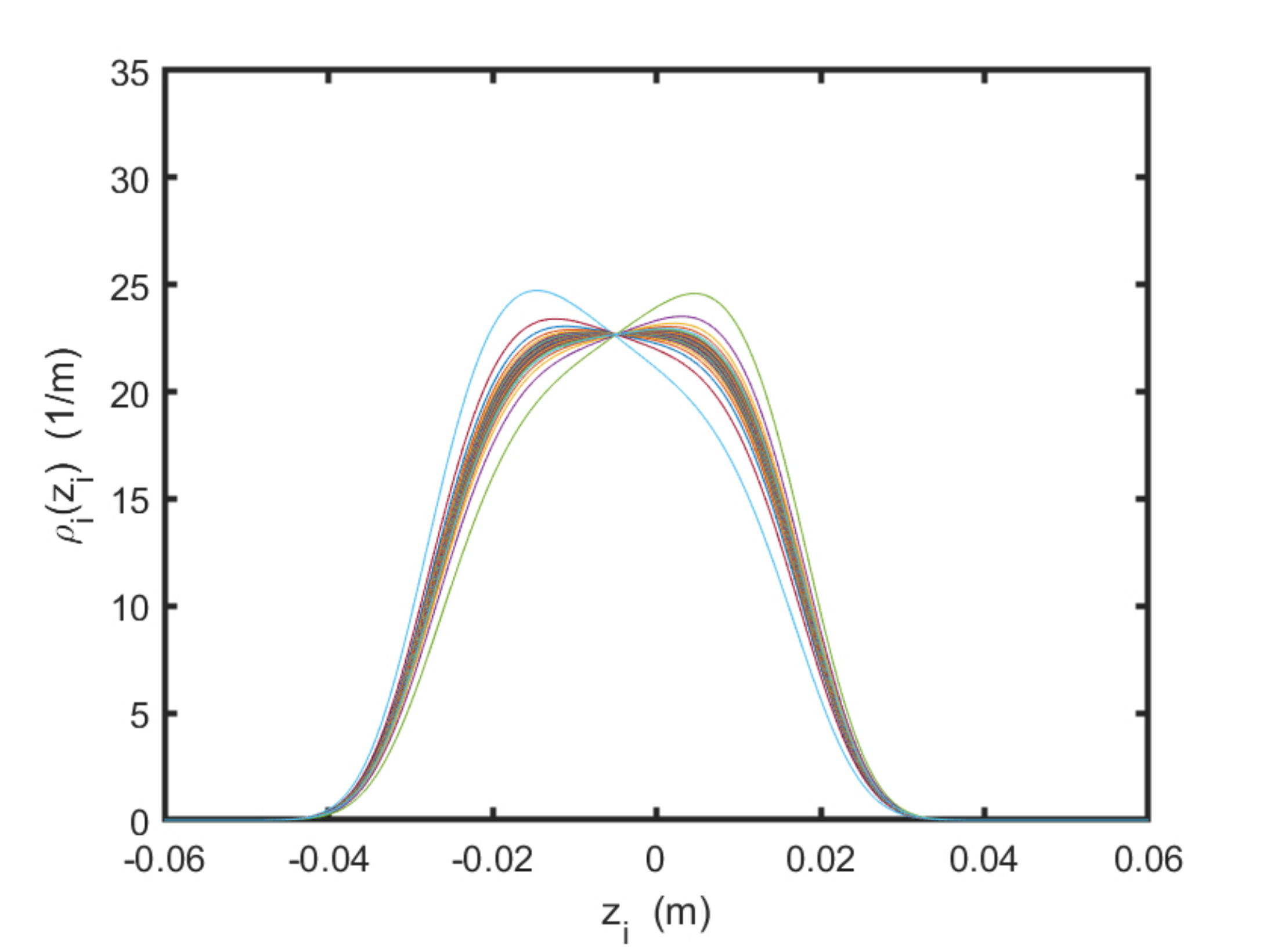}     %rho_284_taper_HHC_I496.pdf
   \caption{Case of tapered bunch charges,\\ MC beam loading omitted, $I_{\rm av}=496$ mA.}
   \label{fig:fig10}
   \end{minipage}
\end{figure}

We next consider the same fill pattern with 11 trains, but with a taper in the bunch charges putting more charge at the ends, according to
a power law as shown in Fig. 15 of \cite{prabI}. This is an example of invoking guard bunches to reduce the effect of gaps. As is seen
in Figs.\ref{fig:fig9} and \ref{fig:fig10}, the guarded inner bunches, which resemble that of the complete fill, are little affected by the MC.
The strong asymmetry between the front and back of the train is perhaps surprising, but it should be noticed that Fig.\ref{fig:fig10} already shows an appreciable front-back asymmetry. The strong amplification of this asymmetry by the MC is in line with its big effects seen generally.

\subsection{Decrease of HHC detuning for over-stretching \label{subsection:smalldf}}
\begin{figure}[htb]
   \centering
   \begin{minipage} [b]{.49\linewidth}
   \includegraphics[width=\linewidth]{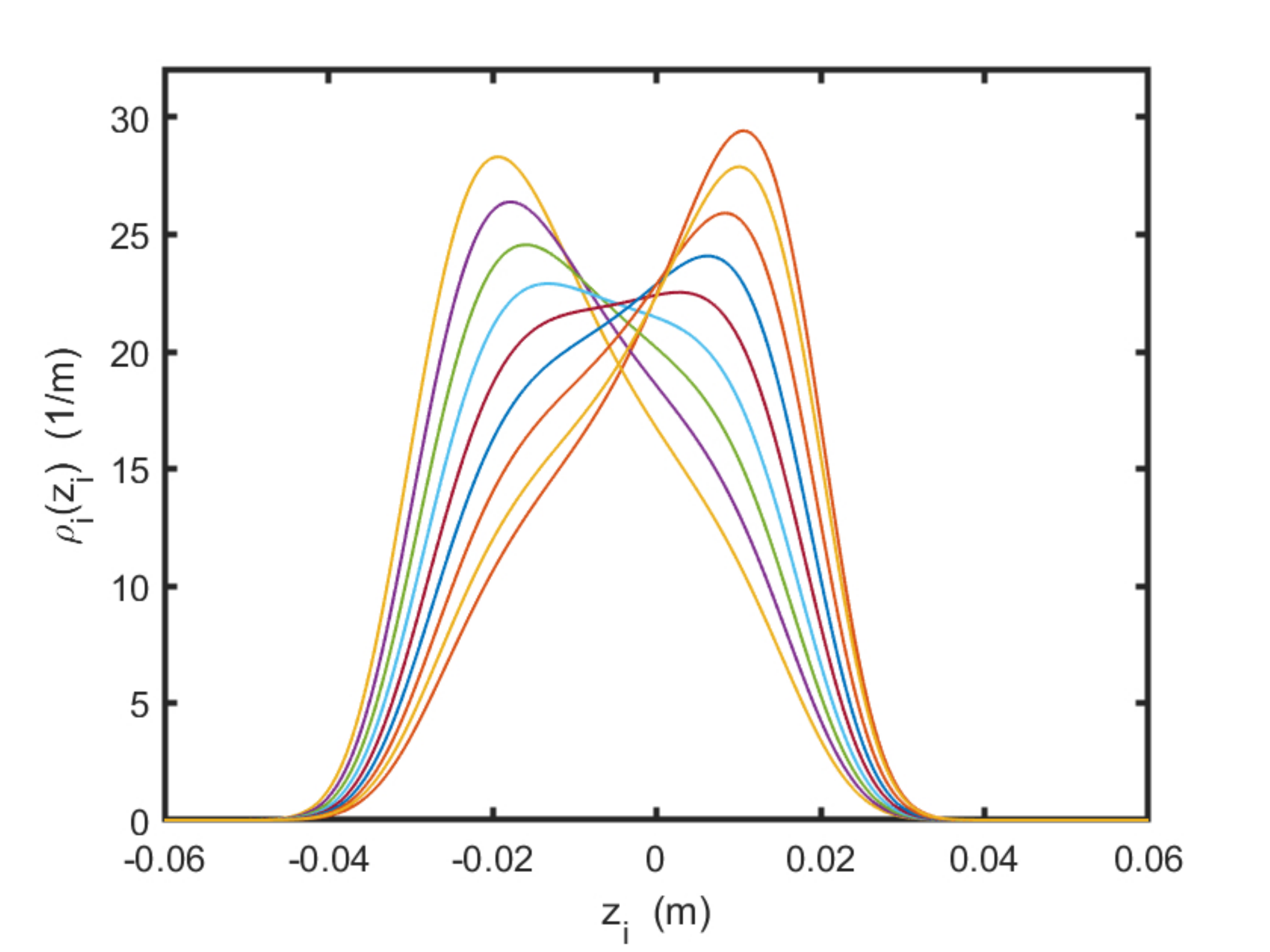}     %rho_284_ng_HHC_MC_df235_I4745.pdf
  \caption{Fill C2 with HHC + MC,\\ detuning $df=235$~kHz, $I_{\rm av}=474.5$~ mA.}
   \label{fig:fig11}
   \end{minipage}
   \begin{minipage} [b]{.49\linewidth}
   \includegraphics[width=\linewidth]{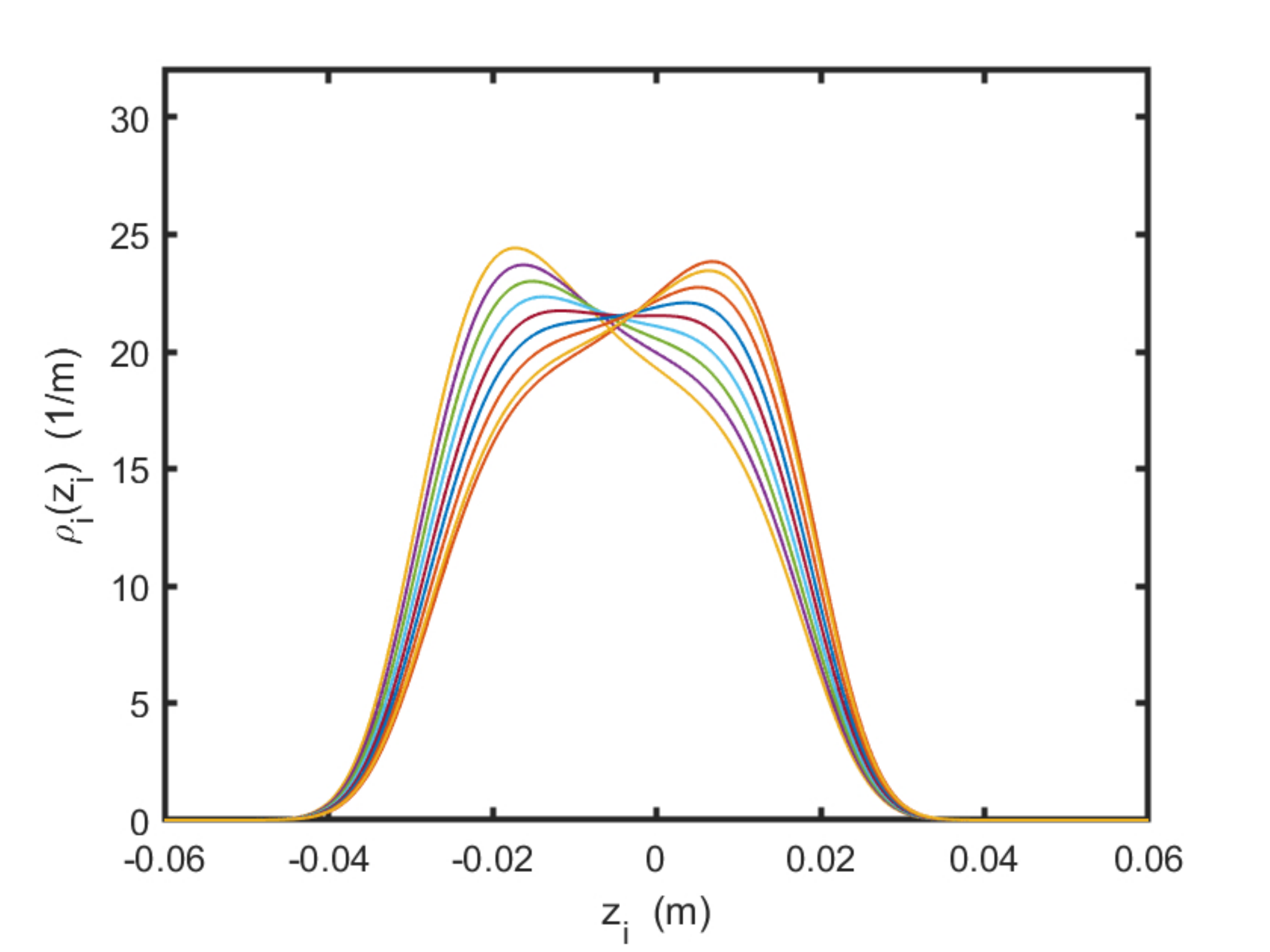}     %rho_284_ng_HHC_df235_I4745.pdf
   \caption{Fill C2 with HHC only,\\ detuning $df=235$~kHz, $I_{\rm av}=474.5$~ mA.}
   \label{fig:fig12}
   \end{minipage}
\end{figure}

There is practical interest in the possibility of over-stretching for an additional increase in the Touschek lifetime. This entails
a decrease in the detuning of the HHC, which produces a larger r.m.s. bunch length but a bunch profile with a dip in the middle, thus a double peak. In our case a decrease from $df=250.2$~kHz to $df=235$~kHz produces a double peak in the model without the MC at full current, as is seen in Fig.4 of \cite{prabI}. We would like to know how this setup looks with the compensated main cavity in play. Not surprisingly, the convergence
of our iterative solution breaks down at a lower current than in the case of the normal detuning; the stronger the bunch distortions the poorer the convergence. With $df=235$~kHz and the MC we can only reach 474.5 mA, which is not enough to see a double peak. Nevertheless it is useful
to compare the result at that current with the result in absence of the MC, as displayed for 9 bunches in a train of 27 in Figs.\ref{fig:fig11} and \ref{fig:fig12}.

Even at a current significantly less that the 500 mA design current the distortion due to the main cavity
is quite large, which leads to the conclusion that {\bf the main cavity must be included in a realistic simulation of over-stretching.}

\subsection{Effect of the short range wake field \label{subsection:short}}
\begin{figure}[htb]
   \centering
   \includegraphics[width=.6\linewidth]{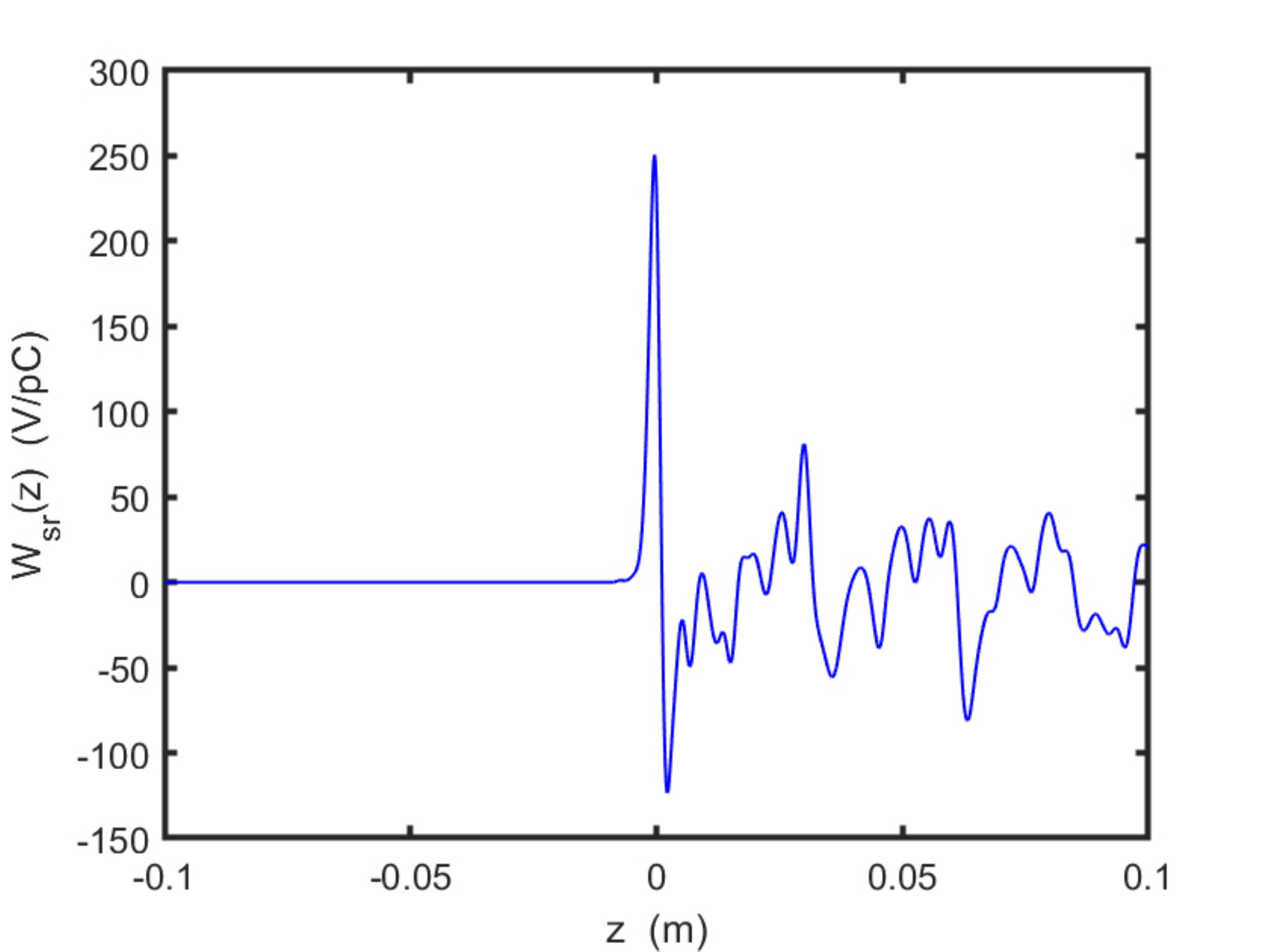}     %wakesr.pdf
  \caption{Wake potential (pseudo - Green function) for the ALS-U storage ring, computed with a 1~mm driving bunch.}
   \label{fig:fig13}
\end{figure}

\begin{figure}[htb]
   \centering
   \includegraphics[width=.6\linewidth]{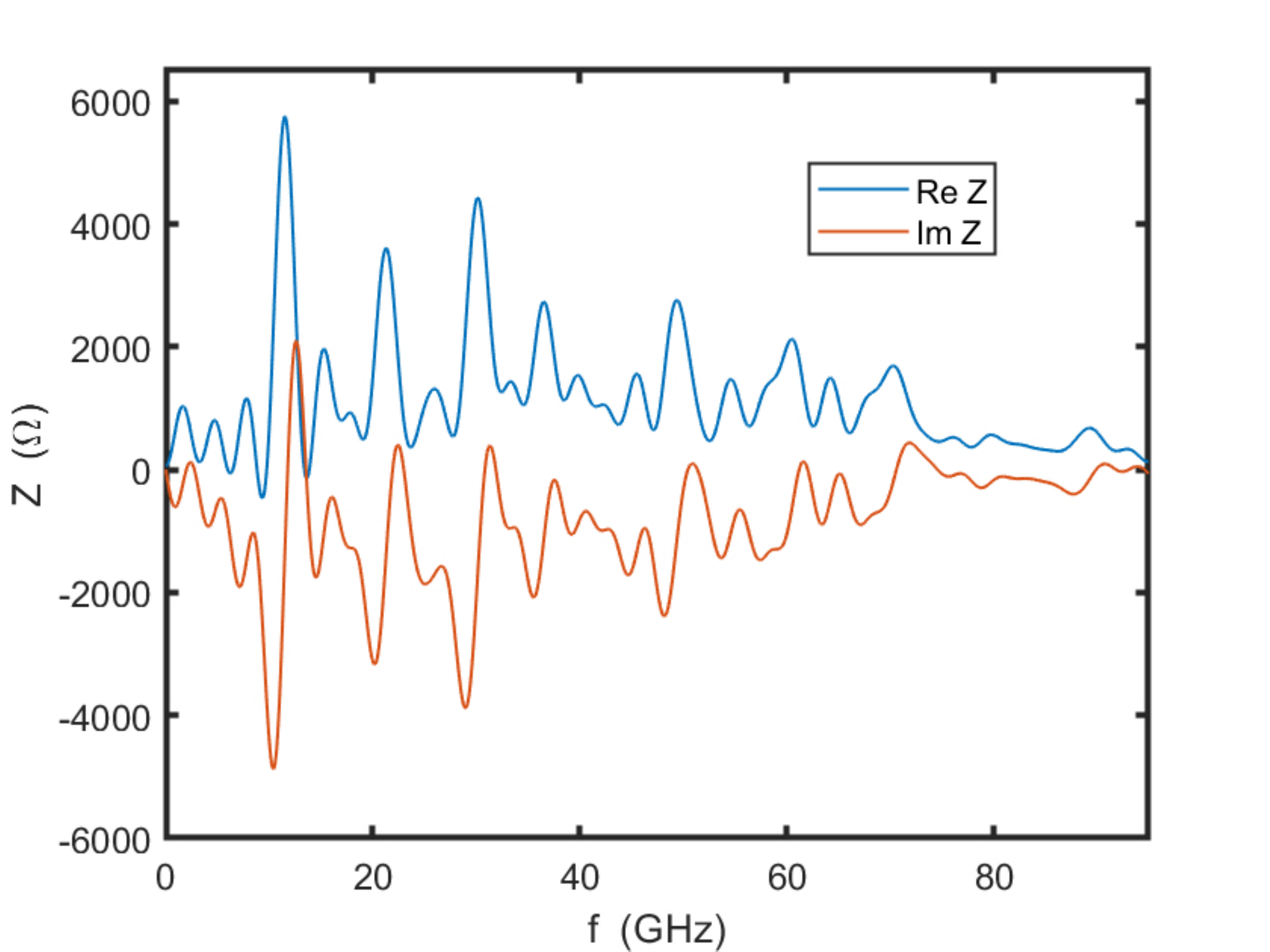}     %z_emp.pdf
  \caption{Longitudinal impedance $Z(f)$ for ALS-U.}
   \label{fig:fig14}
\end{figure}

The short range wake field from various unavoidable corrugations in the vacuum chamber retains importance in the latest storage rings,
in spite of the best efforts to reduce it. Since it can cause substantial bunch distortion in the absence of an HHC, we would like to know how much it affects the operation of the HHC.  A result for the longitudinal wake potential at ALS-U, from a detailed computation by Dan Wang  \cite{dwang},
is shown in Fig.\ref{fig:fig13}.
The corresponding impedance,
\be
Z(f)=\frac{1}{c}\int_{-\infty}^\infty e^{-ikz}W(z)dz\ ,\quad f=kc/2\pi\ , \label{zeedef}
\ee
is plotted  in Fig.\ref{fig:fig14}.

In Ref.\cite{prabI} we suggested that a low-$Q$ resonator wake could be treated on the same footing as the high-$Q$ resonators, and for that
reason we wrote all equations for a general value of $Q$. We recognized, however, that the diagonal term in the potential would now be dominant,
while being nearly negligible in the high-$Q$ case. It could not be treated by the method used in \cite{prabI}, but is easily handled
by the presently adopted method of Section \ref{section:fullsys}.

\begin{figure}[htb]
   \centering
   \begin{minipage} [b]{.49\linewidth}
   \includegraphics[width=\linewidth]{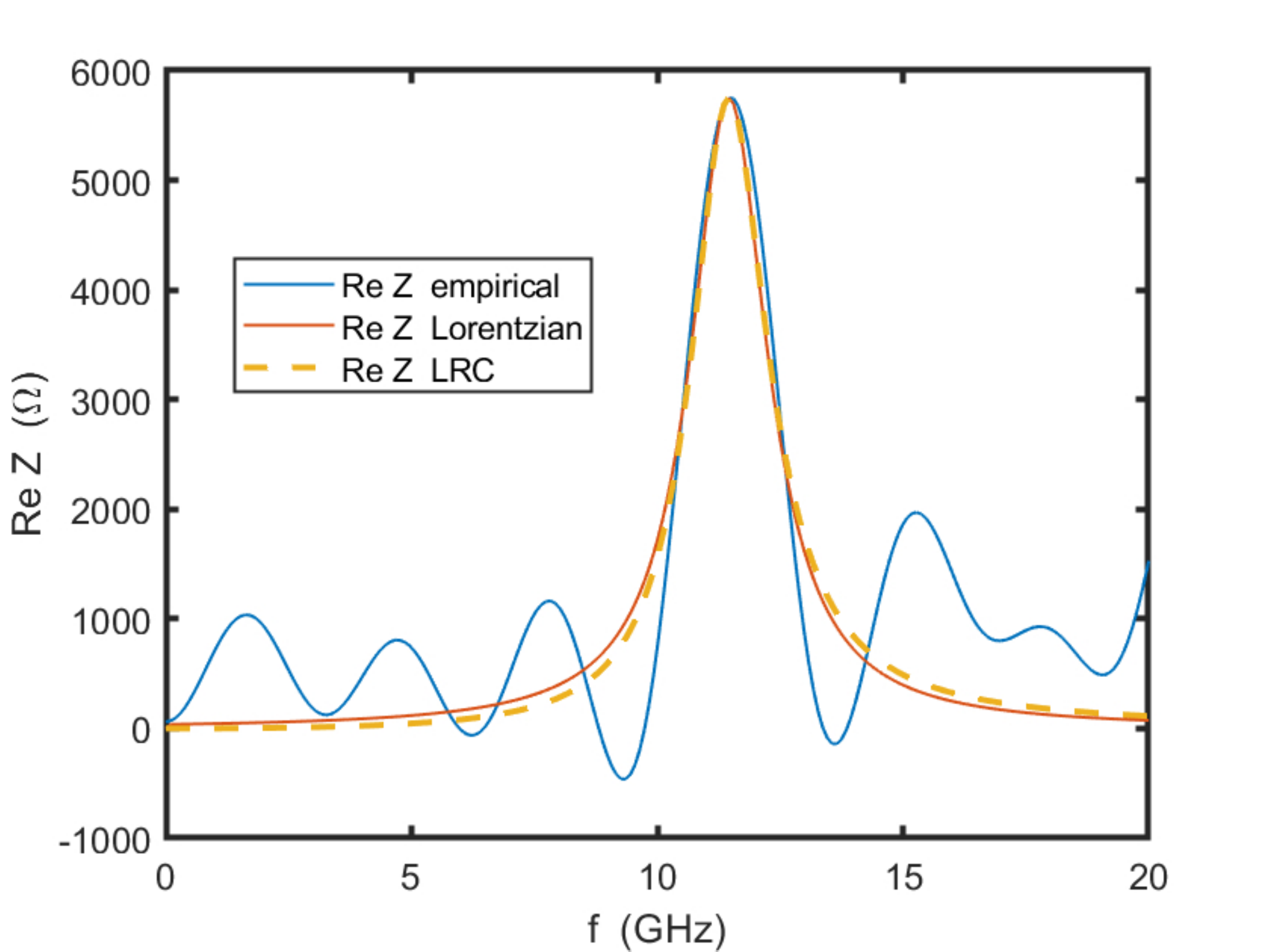}     %rez_fit.pdf
  \caption{Fit of $\rep Z$ to Lorentzian and\\ LRC circuit formulas.}
   \label{fig:fig15}
   \end{minipage}
   \begin{minipage} [b]{.49\linewidth}
   \includegraphics[width=\linewidth]{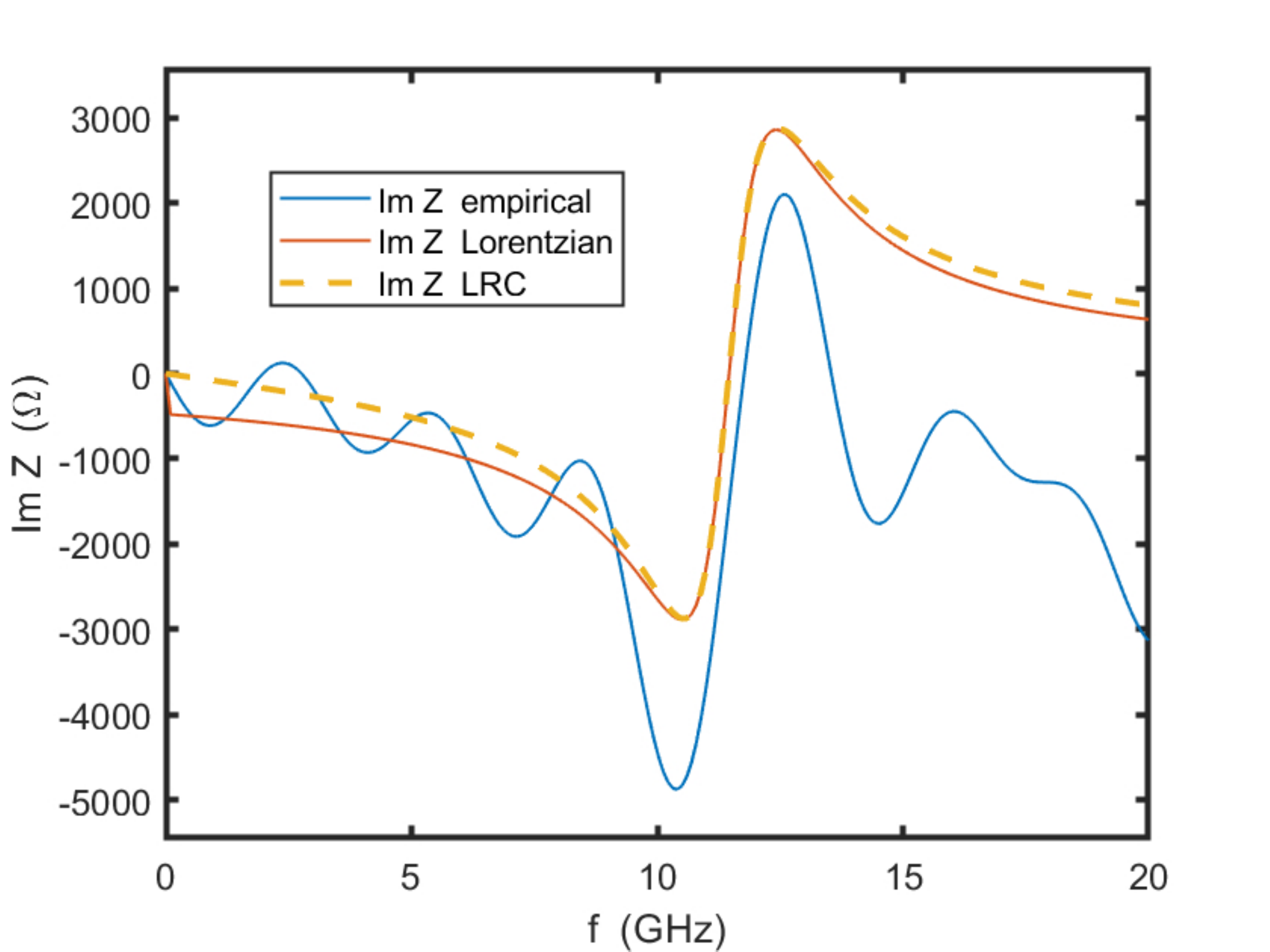}     %imz_fit.pdf
   \caption{Fit of $\imp Z$ to Lorentzian and\\ LRC circuit formulas.}
   \label{fig:fig16}
   \end{minipage}
\end{figure}

For the equilibrium state, the impedance at $f > 20$~GHz is irrelevant, even though it could have a role out of equilibrium. This assertion
follows from the fact that the frequency spectrum of our calculated charge densities never extends beyond 15 GHZ, no matter which wake fields
are included. Consequently, a reasonable step is to concentrate on the first big peak at 11.5 GHz. The wake potential in our equations
(defined in (19) of \cite{prabI})
is based on an impedance as follows, which is of Lorentzian form with half-width $\Gamma/2$:
\be
Z(f)=iR_s\frac{\Gamma}{2}\bigg[\frac{1}{f-f_r+i\Gamma/2}+\frac{1}{f+f_r+i\Gamma/2}\bigg]=Z(-f)^*\ ,\quad \Gamma/2=f_r/2Q \ .\label{zpoles}
\ee
Figures \ref{fig:fig15} and  \ref{fig:fig16} show a fit to (\ref{zpoles}) with parameters as follows:
\be
f_r=11.549~ {\rm GHz}\ ,\quad R_s=5730~ \Omega\ ,\quad Q=6\ .
 \label{zpolespara}
\ee
The fit is rough in the imaginary part, but probably good enough to estimate the magnitude of the effect of the short range wake.

Discussions of low-$Q$ resonator models in the literature usually invoke the impedance of an LRC circuit, $Z(f)=R/(1+iQ(f_r/f-f/f_r))$, often
with $Q$ near 1. As is illustrated in Figures \ref{fig:fig15} and  \ref{fig:fig16}, in our case with $Q=6$ the LRC model does not give a better fit than the simpler Lorentzian, except for enforcing $Z(0)=0$.
At the expense of some complication our equations could be modified to accommodate the LRC form, but that appears to be unnecessary, at least in the present example.

\begin{figure}[htb]
   \centering
   \includegraphics[width=.6\linewidth]{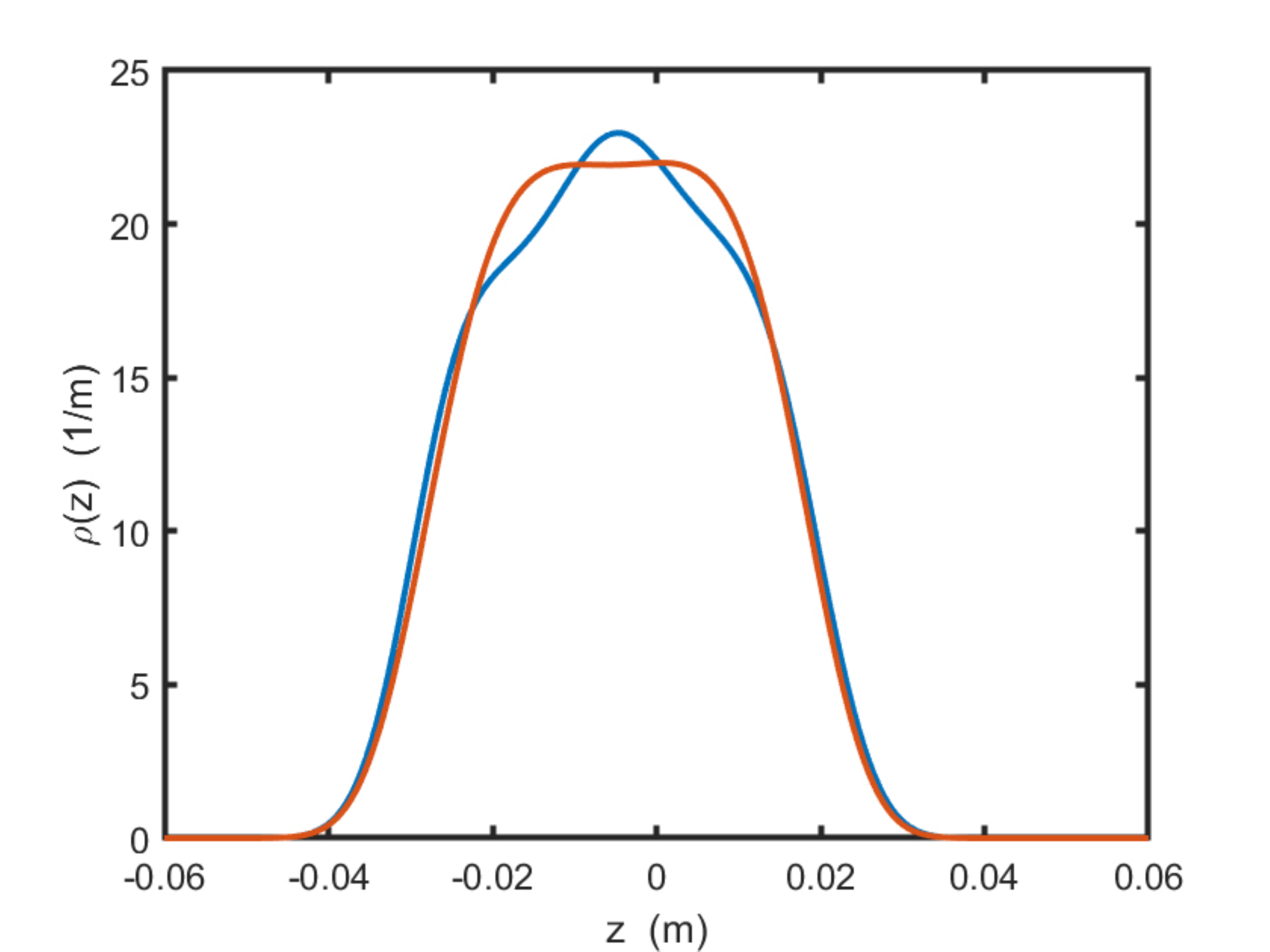}     %rho_328_HHC_SR_I500.pdf
  \caption{Charge density for a complete fill, with HHC plus the first peak in the short range impedance (blue),
  and with HHC alone (red). $I_{\rm av}=500$~mA.}
   \label{fig:fig17}
   \end{figure}

Henceforth, the impedance from (\ref{zpoles}) and (\ref{zpolespara}) will be referred to as SR (short range).
Taking first a complete fill, and including just the HHC and SR, we get the result
of Fig.\ref{fig:fig17}.

\begin{figure}[htb]
   \centering
   \begin{minipage} [b]{.49\linewidth}
   \includegraphics[width=\linewidth]{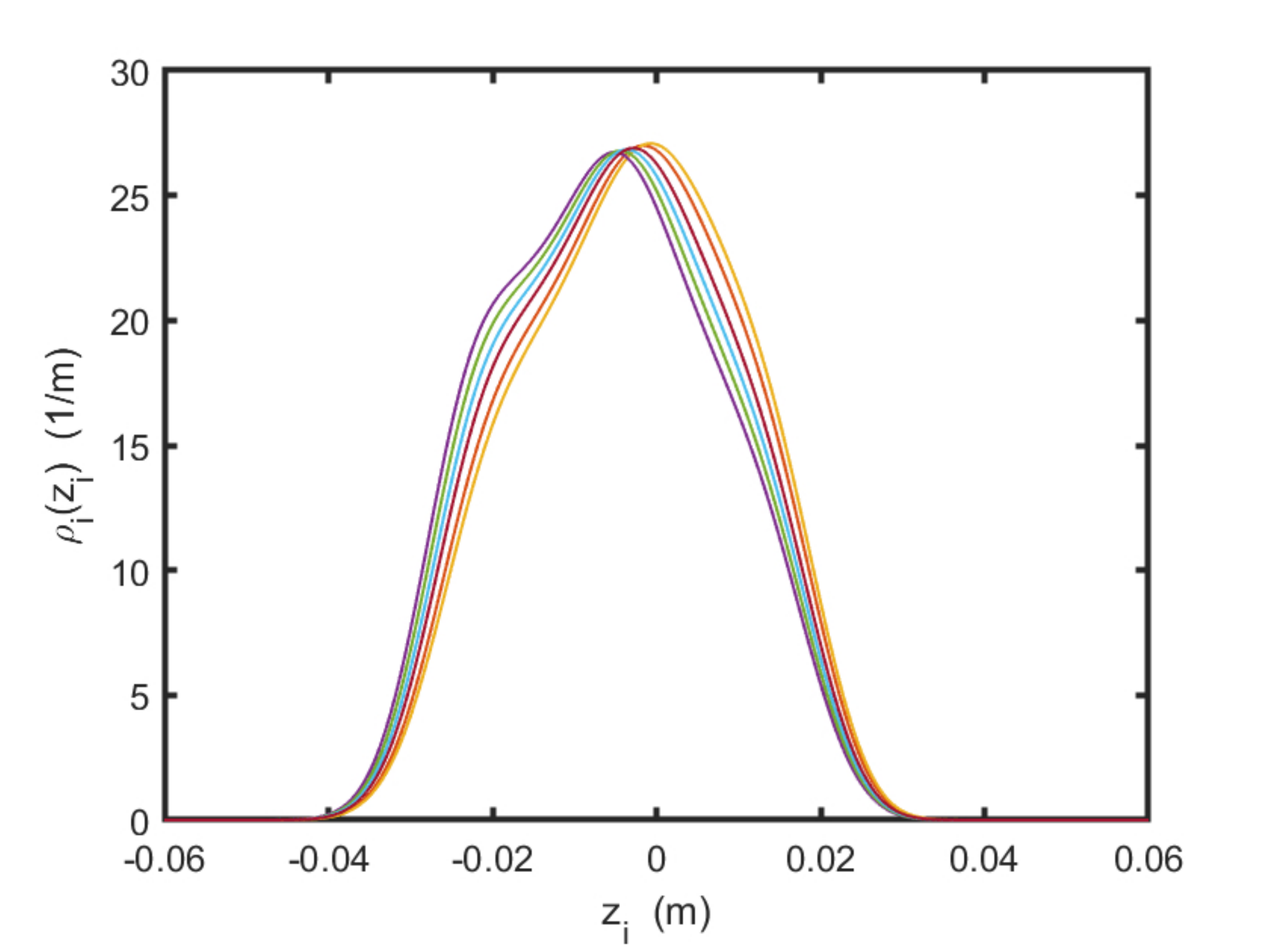}     %rho_284_ng_HHC_SR_I4762.pdf
  \caption{Fill C2, HHC + SR, $I_{\rm av}=476.2$ mA.}
   \label{fig:fig18}
   \end{minipage}
   \begin{minipage} [b]{.49\linewidth}
   \includegraphics[width=\linewidth]{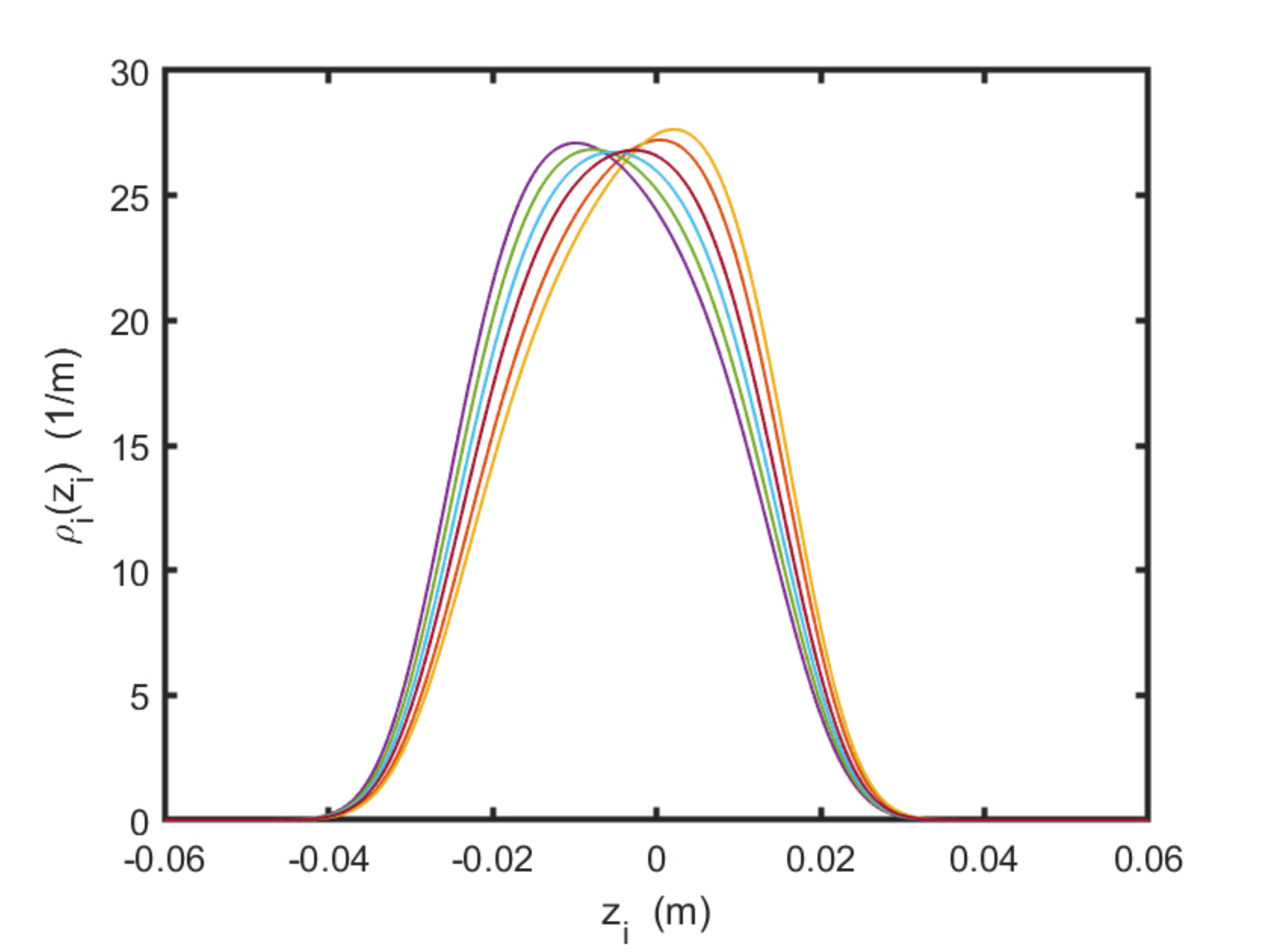}     %rho_284_ng_HHC_I4762.pdf
   \caption{Fill C2, HHC alone, $I_{\rm av}=476.2$ mA.}
   \label{fig:fig19}
   \end{minipage}
\end{figure}

Next we consider the partial fill C2 with distributed gaps as treated in the previous section. Figures \ref{fig:fig18} and \ref{fig:fig19} show the results for HHC+SR and HHC alone.
As expected, the effects of SR are more pronounced in the partial fill than in the complete fill. Correspondingly, the maximum current achieved
is 472.6 mA. As in previous cases we expect a substantially larger effect at the design current of 500 mA.

\subsection{Higher order mode (HOM) of the harmonic cavity \label{subsection:hom}}
\begin{figure}[htb]
   \centering
   \includegraphics[width=.6\linewidth]{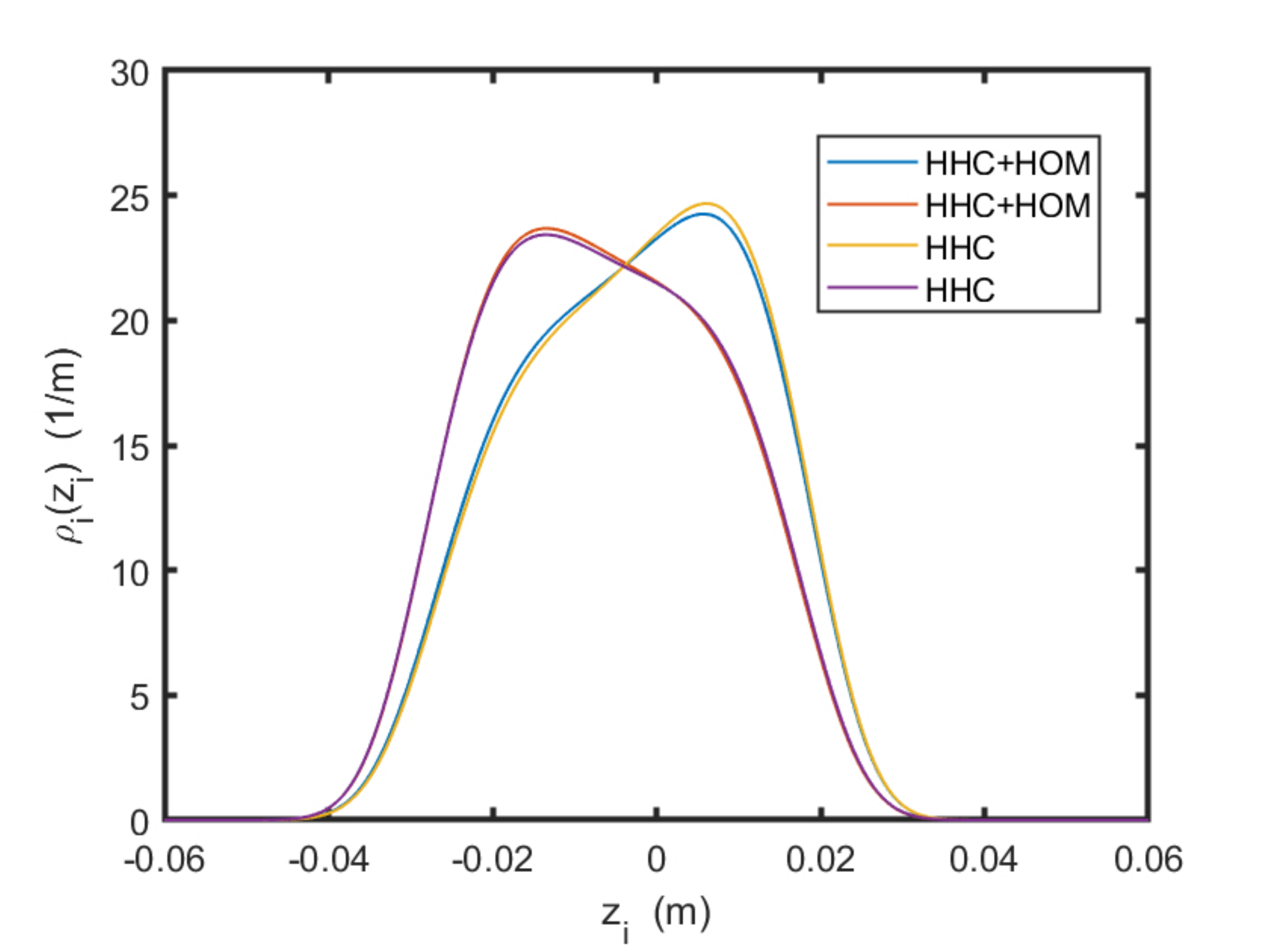}     %rho_284_ng_HHC_HOM_I500.pdf
  \caption{Two bunches in fill C2 with HHC and its higher order mode. $I_{\rm av}=500$ mA. }
   \label{fig:fig20}
   \end{figure}

At the present stage of design the most prominent longitudinal HOM of the HHC for ALS-U is a TM011 mode with the following parameters \cite{luo}:
\be
R_s=3000~ \Omega\ ,\quad Q=80\ ,\quad f_r=2.29~{\rm GHz}   \label{HOMpara}
\ee
A calculation for fill C2 with the HHC and this HOM gave the result of Fig.\ref{fig:fig20}. The effect of the HOM on the charge densities is less than 2\%, in a small shift at the top of the distributions.

At least for the equilibrium state in ALS-U, it appears that the HOM can be neglected. The role of HOM's in longitudinal
coupled-bunch instabilities is discussed in Ref.\cite{cullinan}.

\subsection{The full model: HHC+MC+SR.\label{subsection:full}}
We are now prepared to include the harmonic cavity, the compensated main cavity, and the short range wake, altogether.  The convergence
of the Newton sequence suffers even more than in the previous cases, and the continuation in current reaches only $I_{\rm av}=471.9$~mA.
 Effects seen at this current must severely underestimate what can be expected at 500 mA, because of
the strong variation near the design current that we have observed in every case.

\begin{figure}[htb]
   \centering
   \begin{minipage} [b]{.49\linewidth}
   \includegraphics[width=\linewidth]{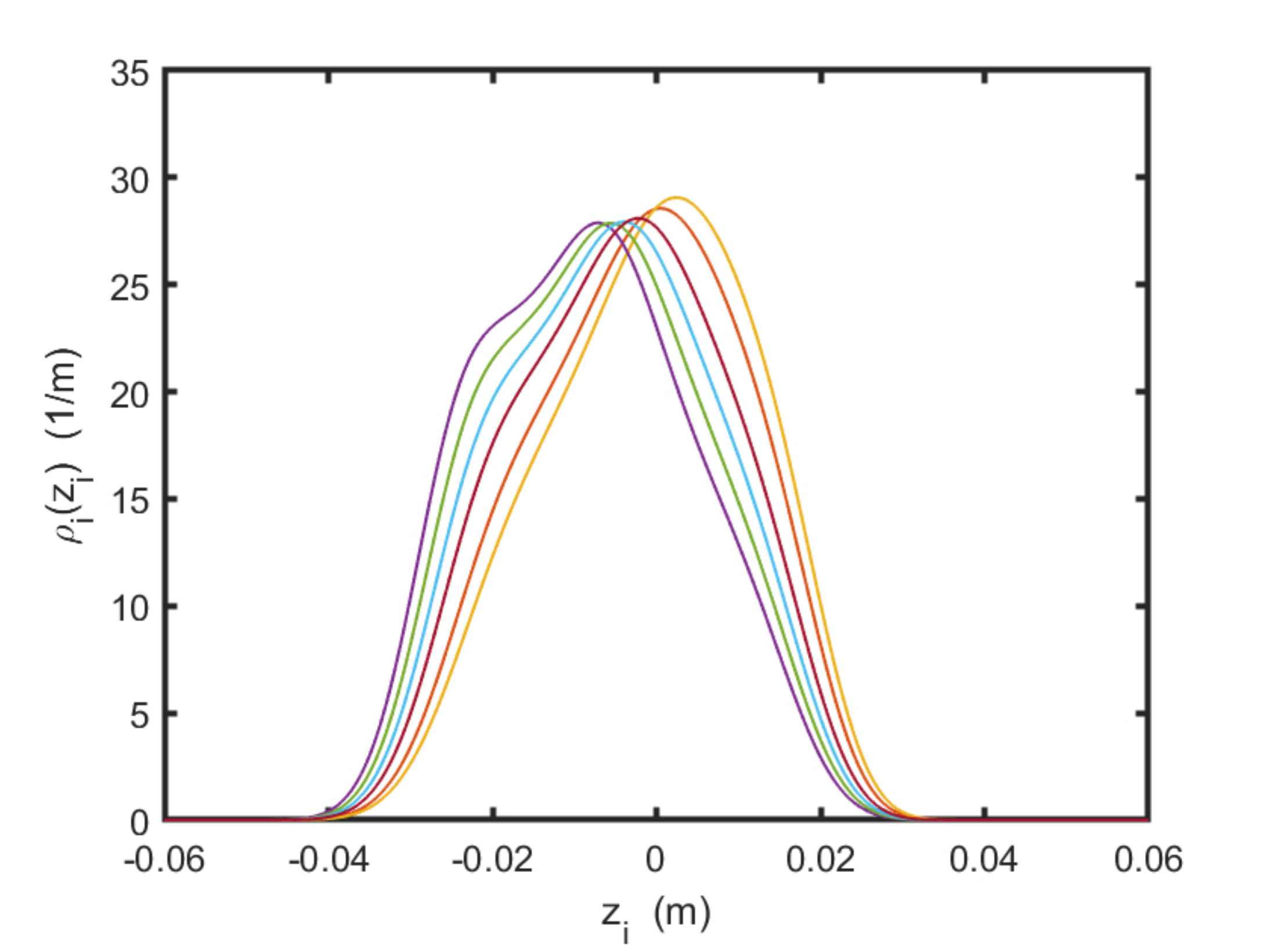}     %rho_284_ng_HHC_MC_SR_I0545.pdf
  \caption{HHC+MC+SR, $I_{\rm av}=471.9$ mA.}
   \label{fig:fig21}
   \end{minipage}
   \begin{minipage} [b]{.49\linewidth}
      \includegraphics[width=\linewidth]{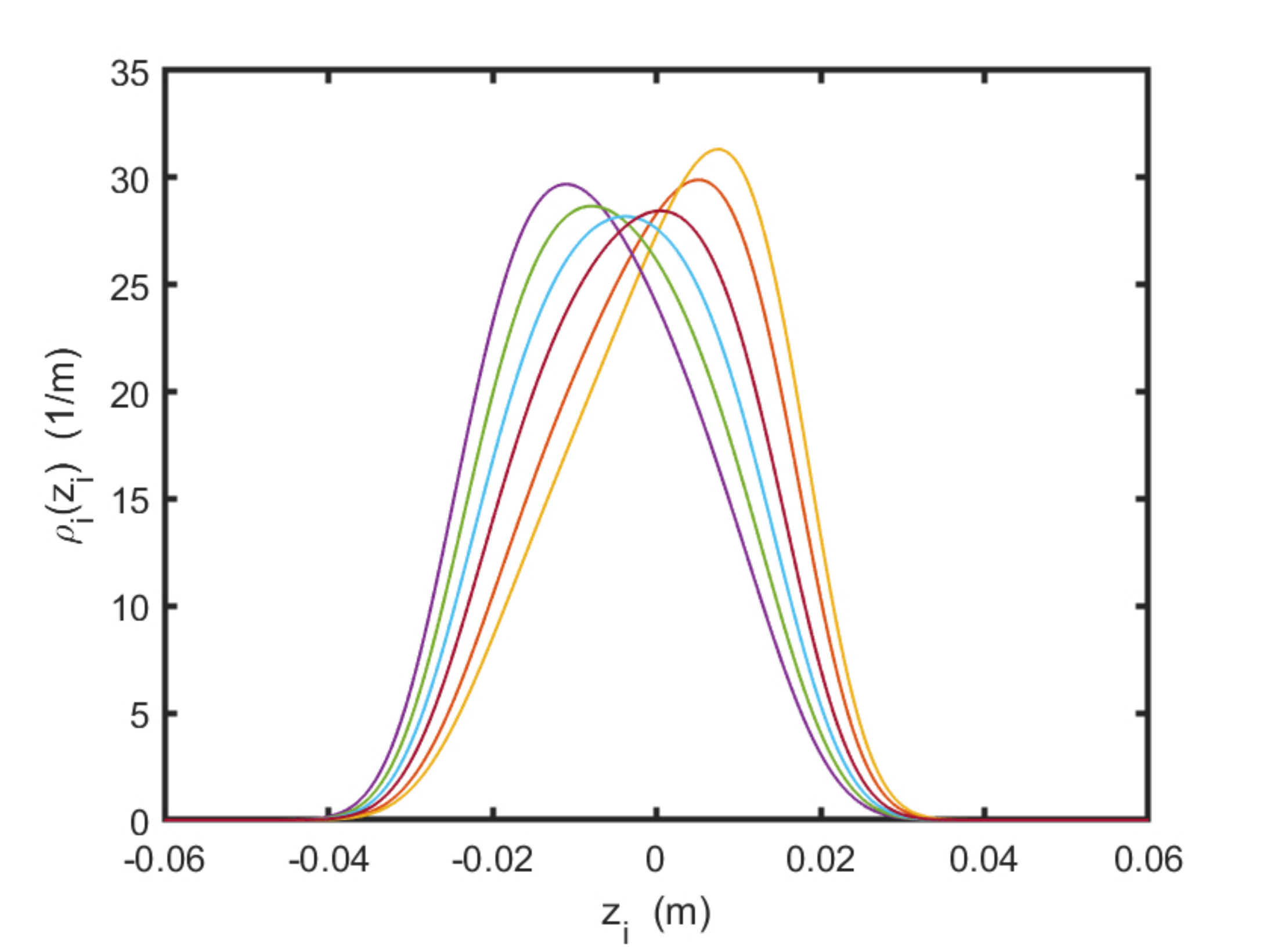}     %rho_284_ng_HHC_MC_I0545.pdf
  \caption{HHC+MC, $I_{\rm av}=471.9$ mA.}
    \label{fig:fig22}
   \end{minipage}
   \end{figure}

For fill C2 we see the charge densities in Figures \ref{fig:fig21} and \ref{fig:fig22}.

\subsection{The case of a single gap, with main cavity beam loading \label{subsection:single}}
\begin{figure}[htb]
   \centering
   \begin{minipage} [b]{.49\linewidth}
   \includegraphics[width=\linewidth]{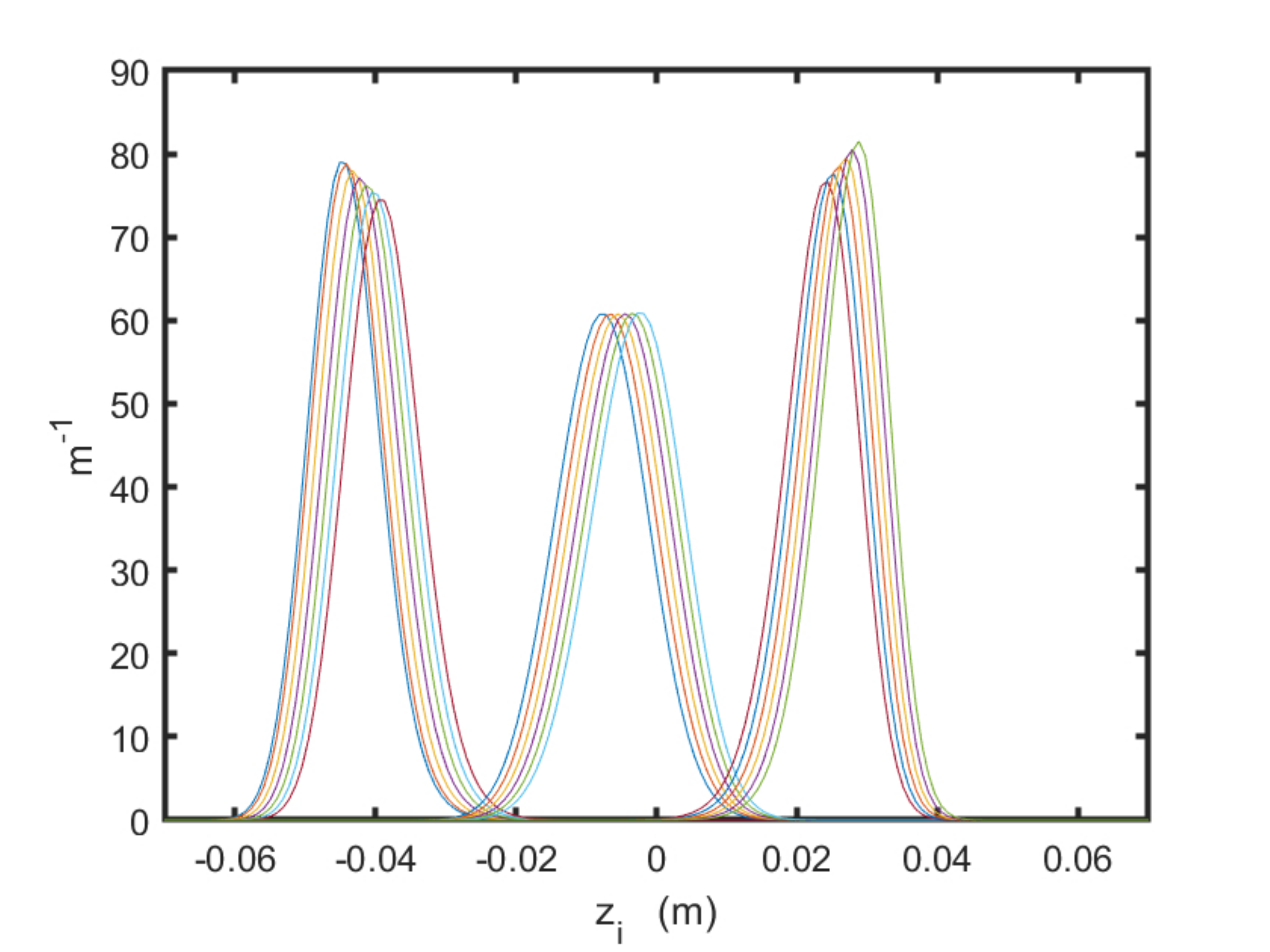}     %rho_284_ng_HHC_MC_SR_I0545.pdf
  \caption{HHC+MC, single gap, $I_{\rm av}=500$ mA.}
   \label{fig:fig23}
   \end{minipage}
   \begin{minipage} [b]{.49\linewidth}
      \includegraphics[width=\linewidth]{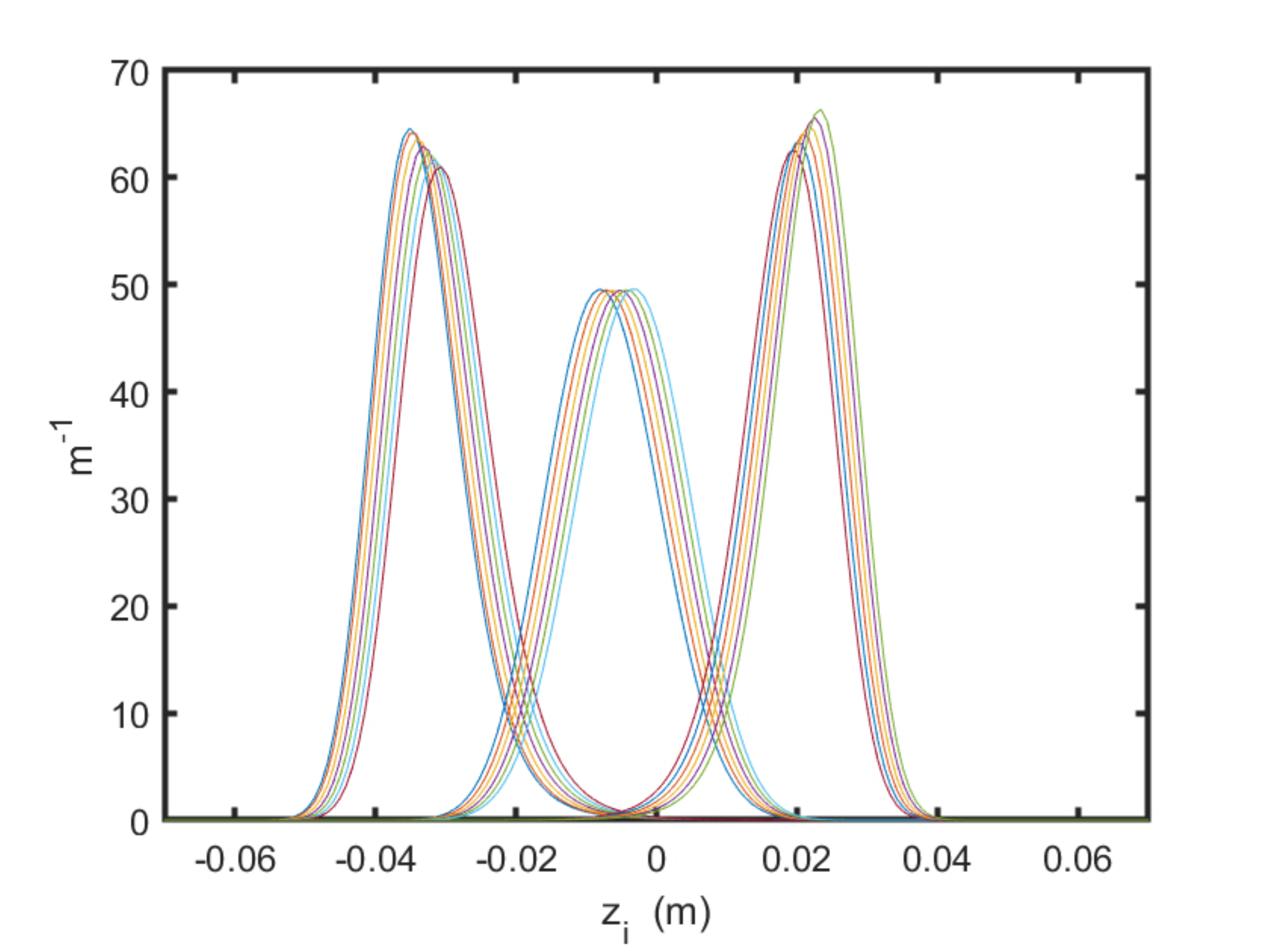}     %rho_284_ng_HHC_MC_I0545.pdf
  \caption{HHC, single gap, $I_{\rm av}=500$ mA.}
    \label{fig:fig24}
   \end{minipage}
   \end{figure}

\begin{figure}[htb]
   \centering
   \includegraphics[width=.6\linewidth]{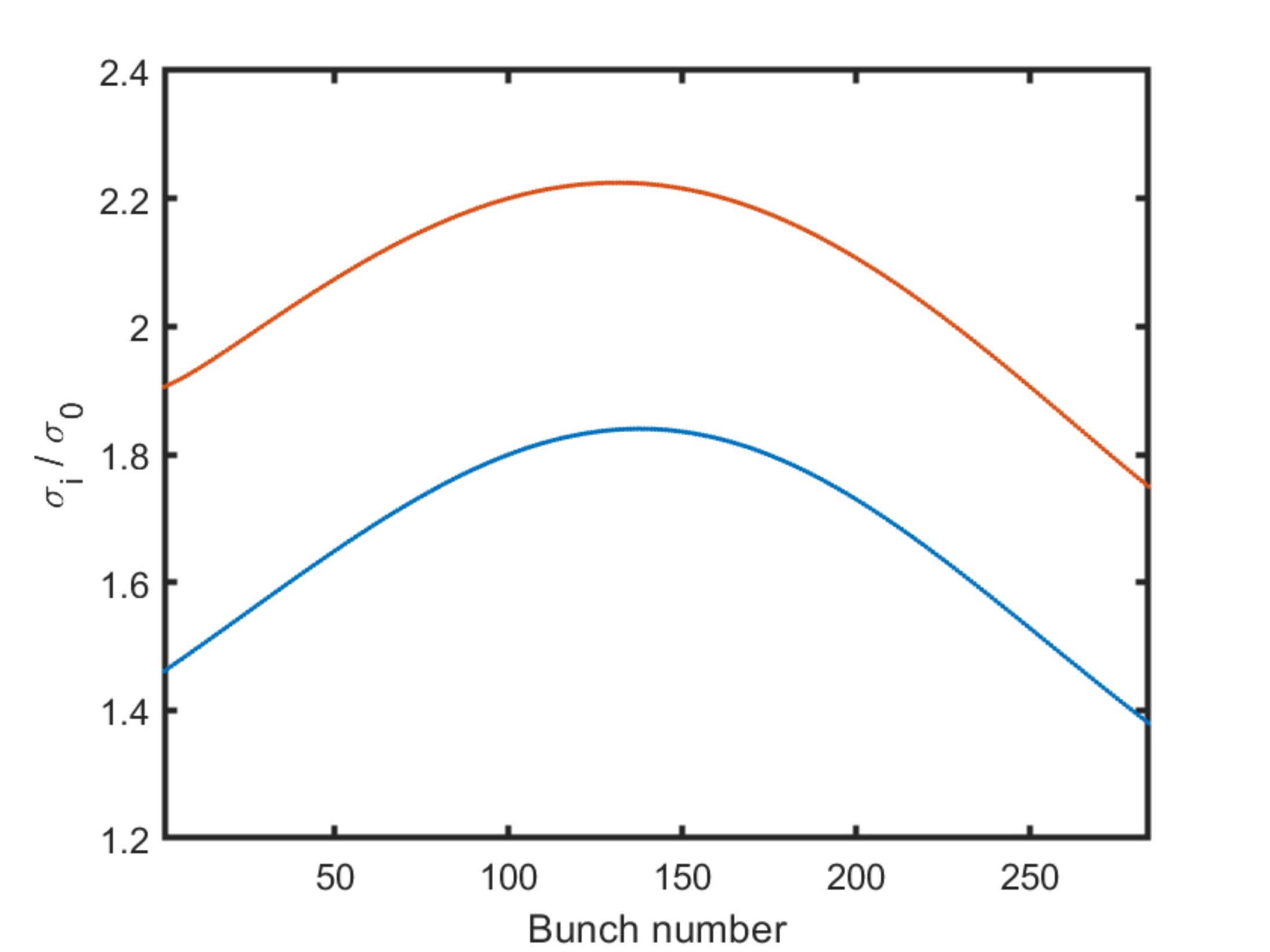}     %rho_284_ng_HHC_HOM_I500.pdf
  \caption{Single gap, bunch lengthening ratio, for HHC+MC (blue) and with HHC alone (red).}
   \label{fig:fig25}
   \end{figure}

It is worthwhile to examine the effect of main cavity beam loading  when there is only a single gap in the fill pattern, even though this
is not directly relevant to the ALS-U design. With 284 bunches, a gap of 44 buckets, and HHC+MC we get the result of Fig.\ref{fig:fig23} for charge densities,
to be compared with the case of HHC alone in Fig.\ref{fig:fig24}. This result could be obtained with the full current of $500$ mA. The graphs show 
6 bunches at the head of the train (right), middle of the train (middle), and end of the train (left).
The bunch lengthening is smaller and the centroid displacement greater when the MC is included. The comparison of bunch lengthenings is shown
in Fig.\ref{fig:fig25}.

\section{Conclusions and outlook  \label{section:outlook}}
Continuing the investigation of Ref.\cite{prabI} we have extended the physical model to include the effect of the main accelerating cavity
in its fundamental mode, previously omitted. We introduced a new algorithm to adjust the parameters of the rf generator voltage so as to compensate the voltage induced
in the cavity by the beam, thus putting the net accelerating voltage at a desired value. When the cavity is excited by a bunch train with gaps this compensation implies a modification of the bunch profiles, which is produced automatically in our scheme.

We illustrated the outcome  for parameters of the forthcoming ALS-U storage ring, revisiting examples treated in \cite{prabI} without the main cavity. The results are similar {\it in modo grosso}, but there are significant quantitative differences, especially in cases of overstretching
of bunches. Generally speaking there is more bunch distortion and less symmetrical patterns in the bunch trains, and the rms bunch lengthening
is a bit smaller and much more variable along the train. Correspondingly, the Touschek lifetime increase is smaller and more variable over a train.

We have not tried to model the feedback system that compensates the beam loading in practice. Our aim was only to show the theoretical existence of an equilibrium state with precise compensation in place.

It was disappointing, and somewhat surprising, to find that the Newton iteration to solve the coupled Ha\"isinski equations encounters
convergence difficulties at large current (near the design current)  when either the main cavity wake or the short range wake is added to the HHC wake. 

A colleague suggested that the failure of convergence might hint at an instability. One should be cautious about such an idea. The
issue here is just the existence of an equilibrium. An equilibrium may or may not be stable under time evolution, so stability is a different issue. 

Our failure to find an equilibrium in some cases of high current may be due to a failure of technique, not necessarily an indication that no equilibrium exists. At high current we are trying to achieve convergence of the Newton iteration close to  a singularity of the Jacobian, but not squarely on the singularity. In this case it is crucial to have a  starting guess sufficiently close to a solution, but in practice the required degree of closeness is unknown. We made some efforts to improve the guess by a seemingly careful continuation in current from the last good solution, but there was no clear success.

A likely remedy for the convergence failure is to return to the conventional formulation of the Ha\"issinski equations as integral equations for the
charge densities, in place of the present formulation as algebraic equations for Fourier amplitudes. For a single Ha\"issinski integral equation discretized  on a mesh in $z$-space, the Newton iterative solution is ultra-robust, converging at currents far beyond realistic values \cite{bobkarl}. It seems likely that similar good behavior will hold for the coupled integral equations. The size of the discretized system does not grow with the number of resonator wakes, in contrast to the present system, and the full $z$-space description of the short range wake could be invoked in place of the low-$Q$ resonator model.

To make this $z$-space formulation feasible on modest computer resources we can assume that all bunch sub-trains are identical, and all separated
by identical gaps. For the ALS-U this would mean artificially increasing the harmonic number from 328 to 330, and having 11 trains of 26 separated
by gaps of 4 buckets. Then we have 26 independent charge densities, which can adequately be described by 100 mesh points each. Thus the Jacobian of the Newton iteration is $2600\times 2600$,  a modest size that will yield a very quick computation.

Moreover, this identical train model would make it feasible to do a time-domain solution of the coupled Vlasov-Fokker-Planck equations by
the method of local characteristics (discretizing the Perron-Frobenius operator) \cite{senigallia}. This could answer the urgent question of
stability of the equilibria, and provide a window to the dynamics out of equilibrium. Incidentally, determination of the instability threshold by the linearized Vlasov system could also be attempted.

It was gratifying to find that the sub-iteration to enforce the main cavity compensation converged very quickly whenever the main iteration converged. It can be employed in the same way in the proposed $z$-space system. Also, the formalism for multiple resonator wakes will still be advantageous in the $z$-space scheme.

\section{Acknowledgments \label{section:acknow}}
I thank Teresia Olsson for a helpful correspondence, Dan Wang for her wake potential, and Tianhuan Luo for information on the HHC design.
Marco Venturini posed the main cavity compensation problem in general terms. Karl Bane encouraged the study of the short range wake. This work was supported in part by the U. S. Department of Energy, Contract Nos. DE-AC03-76SF00515. My work is aided by an affiliation with Lawrence Berkeley National Laboratory as Guest Senior Scientist.

\appendix

\section{Diagonal terms in the potential. \label{section:appA}}
Here we find the formula for a generic term in the first sum of (\ref{udef2}). For this we revert to
the notation used in the case of a single resonator.

The term in question is the last term of (51) in \cite{prabI}, defined through (60) of that paper, as follows:
\bea
&&U_i^d(z_i)=\frac{e^2N\omega_rR_s\eta\xi_i}{Q}\bigg[\int_0^{z_i}d\zeta\int_{-\Sigma}^{\zeta}\exp(-k_r(\zeta-u)/2Q)\cos(k_r(\zeta-u)+\psi)\rho_i(u)du\nonumber\\
&&+\int_0^{z_i}d\zeta\int_{\zeta}^\Sigma\exp(-k_r(\zeta-u+C)/2Q)\cos(k_r(\zeta-u+C)+\psi)\rho_i(u)du\bigg]\ . \label{vdm}
\eea
The repeated integrals can be replaced by single integrals through integration by parts. First apply the
double angle formula to the cosine, so as to bring out factors of $\cos(k_ru)$ and $\sin(k_ru)$. The $u$-integrals
involving those factors are functions of $\zeta$, which are to be differentiated in the partial integration with respect to
$\zeta$. The corresponding integration with respect to $\zeta$ is done with the help of (55) and (56) (as indefinite integrals) in \cite{prabI}. The result is
\bea
&&U^d_i(z_i)=\frac{ce^2NR_s\eta\xi_i}{Q(1+(1/2Q)^2)}\big[~I_1+I_2~\big]\ ,\nonumber\\
&&I_1=\int_{-\Sigma}^{z_i}\exp(k_ru/2Q)\bigg[a(z_i)\cos(k_ru)
+b(z_i)\sin(k_ru)\bigg]\rho_i(u)du\nonumber\\
&&\hskip .8cm -\big(\sin\psi-\frac{1}{2Q}\cos\psi\big)\int_{-\Sigma}^{z_i}\rho_i(u)du\ ,\nonumber\\
&&I_2=\int^{\Sigma}_{z_i}\exp(k_ru/2Q)\bigg[a(z_i+C)\cos(k_ru)
+b(z_i+C)\sin(k_ru)\bigg]\rho_i(u)du\nonumber\\
&&\hskip .8cm+\exp(-k_rC/2Q)\big(\sin(k_rC+\psi)-\frac{1}{2Q}\cos(k_rC+\psi)\big)\int_{-\Sigma}^{z_i}\rho_i(u)du\ , \nonumber\\
&& a(z)=\exp\big(-k_rz/2Q\big)\big(\sin(k_rz+\psi)-\frac{1}{2Q}\cos(k_rz+\psi)\big)\ ,\nonumber\\&&b(z)=-\exp\big(-k_rz/2Q\big)\big(\cos(k_rz+\psi)+\frac{1}{2Q}\sin(k_rz+\psi)\big) \ . \label{ui}
\eea
Here we have dropped and added terms independent of $z_i$, which only affect the normalization (\ref{Adef}), and
have used the double angle formula in reverse to consolidate some terms. Writing $\int_{z_i}^\Sigma=\int_{-\Sigma}^\Sigma-\int_{-\Sigma}^{z_i}$,
we see that there are three different integrals to evaluate,
\be
\int_{-\Sigma}^{z_i}\big[~1,~\exp(k_ru/2Q)\cos(k_ru),~\exp(k_ru/2Q)\sin(k_ru)~\big]\rho_i(u)du\ ,   \label{intfu}
\ee
which can be built up stepwise on a mesh in $z_i$. Thus we can compute and store the diagonal terms on the mesh in negligible time.
Note that $I_2$ is totally negligible for the small $Q$ that we encounter in representing the geometric wake, owing to the tiny
prefactor $\exp(-k_rC/2Q)$.

Summing (\ref{ui}) over the $n_r$ choices of the resonator parameters $k_r, R_s, Q, \eta, \psi$ we obtain the first term of (\ref{udef2}).

\end{document}